\documentclass[smallextended]{svjour3}
\smartqed
\usepackage{graphicx}
\usepackage{booktabs}
\usepackage{tabularx}
\usepackage{array}
\usepackage{amsmath,amssymb}
\usepackage[authoryear,round]{natbib}
\usepackage{url}
\usepackage{textcomp}
\usepackage{xcolor}
\usepackage{listings}
\usepackage[T1]{fontenc}
\usepackage[utf8]{inputenc}
\journalname{Empirical Software Engineering}
\begin{document}

\title{VeeReview: Measuring Build--Verify Pairs in Agent-Authored Pull Requests from a V-Model Perspective}
\titlerunning{VeeReview: Build--Verify Pairs in Agent-Authored Pull Requests}
\author{Gabriel Dautovi \and Jie JW Wu}
\authorrunning{G. Dautovi and J. JW Wu}
\institute{Gabriel Dautovi (corresponding author) \at Department of Computer Science, Michigan Technological University, Houghton, Michigan, United States \\ \email{gdautovi@mtu.edu}
\and Jie JW Wu \at Department of Computer Science, Michigan Technological University, Houghton, Michigan, United States \\ \email{jie.jw.wu@mtu.edu}}
\date{Received: date / Accepted: date}

\maketitle

\begin{abstract}
\textit{Context:} Reviewing a pull request requires verification evidence that addresses the proposed change. Coding agent benchmarks and studies of pull request outcomes do not directly assess this relationship. \textit{Objective:} We test whether the V-Model's build and verify pairs can measure verification evidence in pull requests and support comparisons between agent authored and human-authored changes. \textit{Method:} VeeReview classifies four build and verify pairs using repository signals and a language-model judge. Three pairs use model judgements constrained by deterministic signals, and a rule over the diff and repository tree decides Module Design $\leftrightarrow$ Unit Test. We evaluated VeeReview against a 25-pull request reference set labelled by two raters and applied it to 1,454 matched pairs of agent and human pull requests from 741 repositories. \textit{Results:} VeeReview assigned lower build-only rates to agent-authored pull requests on all four pairs, contrary to the preregistered hypothesis (mixed-effects odds ratios 0.52 to 0.86). Architecture $\leftrightarrow$ Integration was not significant. Standardization for linked issues, shipped tests and size reduced matched-rate differences by 34\% to 60\%, and the rule-decided pair's standardized interval included zero. The instrument did not meet its preregistered calibration criterion: agreement with adjudicated labels was 4\% to 60\% against a required 80\%, and inter-rater agreement was weak ($\kappa$ = 0.154). \textit{Conclusions:} The rule-decided pair shows that a process measure judged by a language model partly reflects how a pull request is presented rather than the verification work behind it. We report the cohort differences as observations made with this instrument, and we argue that such measures need a mechanically decided component and a validated human reference before they can support comparisons of verification practice.
\end{abstract}

\section{Introduction}

Before integrating a pull request, a reviewer must decide whether the proposed change has been adequately verified. Passing tests are only part of that decision. The evidence must also address the change's requirements, system behavior, interfaces and modified logic. Autonomous coding agents make this question increasingly relevant. The AIDev dataset contains 932,791 agent-authored pull requests from five named agents across 116,211 repositories (Li et al. 2026). As agents take on issues, modify code and submit pull requests, maintainers must assess both the change and its supporting evidence (Li et al. 2025).

Existing evaluations address only part of this review task. Task-resolution benchmarks test whether an agent can solve a specified problem, often by-passing tests on curated tasks (Jimenez et al. 2024). Pull-request studies examine whether changes are merged, closed or revised (Abujadallah et al. 2026), but those outcomes also depend on reviewer attention, repository policy and contributor relationships. Neither approach directly assesses whether the submitted verification evidence addresses the change. It remains unclear whether pull-request artifacts can support a reliable measure of this relationship and, in turn, a comparison of agent-authored and human-authored work.

The V-Model offers a way to organize this evidence. It pairs development and verification activities at corresponding levels of abstraction (Forsberg and Mooz 1991; Boehm 1984). For a pull request, a change to module logic calls for relevant unit-test evidence, while an interface change calls for integration-test evidence. These distinctions can tell a reviewer more than a count of added test files. We use the pairing principle to classify evidence. We do not assume that repositories follow the V-Model lifecycle or that evidence missing from a pull request means no verification took place.

We developed VeeReview to classify four build and verify pairs from pull-request artifacts. It extracts deterministic repository signals and uses a language model to judge three pairs that require interpretation. A separate rule decides Module Design $\leftrightarrow$ Unit Test from changed files and repository-tree heuristics, independently of the model's verdict. This design lets us compare model-judged states with a model-independent procedure.

We applied VeeReview to agent-authored pull requests sampled from AIDev and to human-authored pull requests matched within the same repositories on time and approximate size. Matching reduces some differences in repository context and timing but does not make the cohorts interchangeable. The preregistered study addresses three research questions:
\begin{itemize}
\item \textbf{RQ1 (Prevalence).} How do agent-authored and human-authored pull requests differ in the match state of each build and verify pair?
\item \textbf{RQ2 (Locus).} Which pair do agents most often leave build-only, and does that vary by pull request type and by authoring agent?
\item \textbf{RQ3 (Size).} How does pull request size relate to matching, and does the relationship differ by agent?
\end{itemize}

The preregistered RQ1 hypothesis predicted higher build-only rates for agent-authored pull requests, especially at Module Design $\leftrightarrow$ Unit Test. VeeReview instead assigned lower rates to the agent cohort on all four pairs, by 3.3 to 13.3 percentage points. The preregistered mixed-effects analysis found a difference on three pairs, but not on Architecture $\leftrightarrow$ Integration. The cohorts also differed in issue linking and test-file changes. Standardizing for these characteristics and pull-request size reduced the matched-rate gaps by 34\% to 60\%, and the standardized interval for the rule-decided pair included zero. These are conditional descriptive associations. They do not show that pull-request presentation caused the differences.

Instrument validity qualifies how these results should be read. Our preregistered criterion required at least 80\% agreement with an independent human-labelled reference on every pair. Agreement with the adjudicated labels ranged from 4\% to 60\%, and the two raters themselves agreed weakly ($\kappa$ = 0.154). This reference cannot reliably establish instrument accuracy or differences in error between cohorts. Relative to the reference labels, the instrument withheld matched credit more often than it awarded it, but that pattern does not show that both cohorts were measured with equal error. We therefore report the cohort findings as observations made with this instrument rather than as confirmed differences in verification practice.

The paper makes four contributions to the empirical study of verification evidence in pull-based development.
\begin{itemize}
\item An inspectable build and verify classification procedure. VeeReview exposes its deterministic signals, model judgements, rule-based overrides and evidence checks (Section 3). We release the code and configuration so that others can reproduce the procedure and validate it further, while recognizing that individual model-generated verdicts can change across identical runs.
\item A matched comparison of agent-authored and human-authored pull requests. We analyze 1,454 pairs from 741 repositories, fit the preregistered model and document departures from the preregistration (Sections 4 and 5).
\item A sensitivity analysis that separates observable presentation from measured verification. We standardize cohort differences for issue linking, test shipping and size, then compare the model-judged pairs with the rule-decided proxy for Module Design $\leftrightarrow$ Unit Test (Section 5.2). This contrast shows how much of a model-judged process measure can be attributed to what the judge reads.
\item An evaluation of the instrument's repeatability and agreement with human raters, reported in full. The instrument did not meet the preregistered calibration threshold, and individual verdicts are unstable across reruns (Section 5.1). These results qualify the cohort comparison and motivate the validation steps we outline.
\end{itemize}

Section 2 introduces the V-Model's pairing principle and the AIDev dataset. Section 3 describes VeeReview, and Section 4 explains the study design and instrument evaluation. Section 5 reports the results. Section 6 discusses their implications, Section 7 examines threats to validity, Section 8 reviews related work and Section 9 concludes.

\section{Background}

\subsection{The V-Model and its pairing claim}

The V-Model pairs development and verification activities in software and systems engineering (Forsberg and Mooz 1991; INCOSE 2015). Figure 1 shows its conventional form: requirements, system design, architecture and module design correspond to acceptance, system, integration and unit testing. Although often presented as a lifecycle, the feature relevant here is the relationship between work at a particular level of abstraction and the evidence that checks it. This relationship defines the four pair categories in our study.

\begin{figure}[!htbp]\centering
\includegraphics[width=\linewidth]{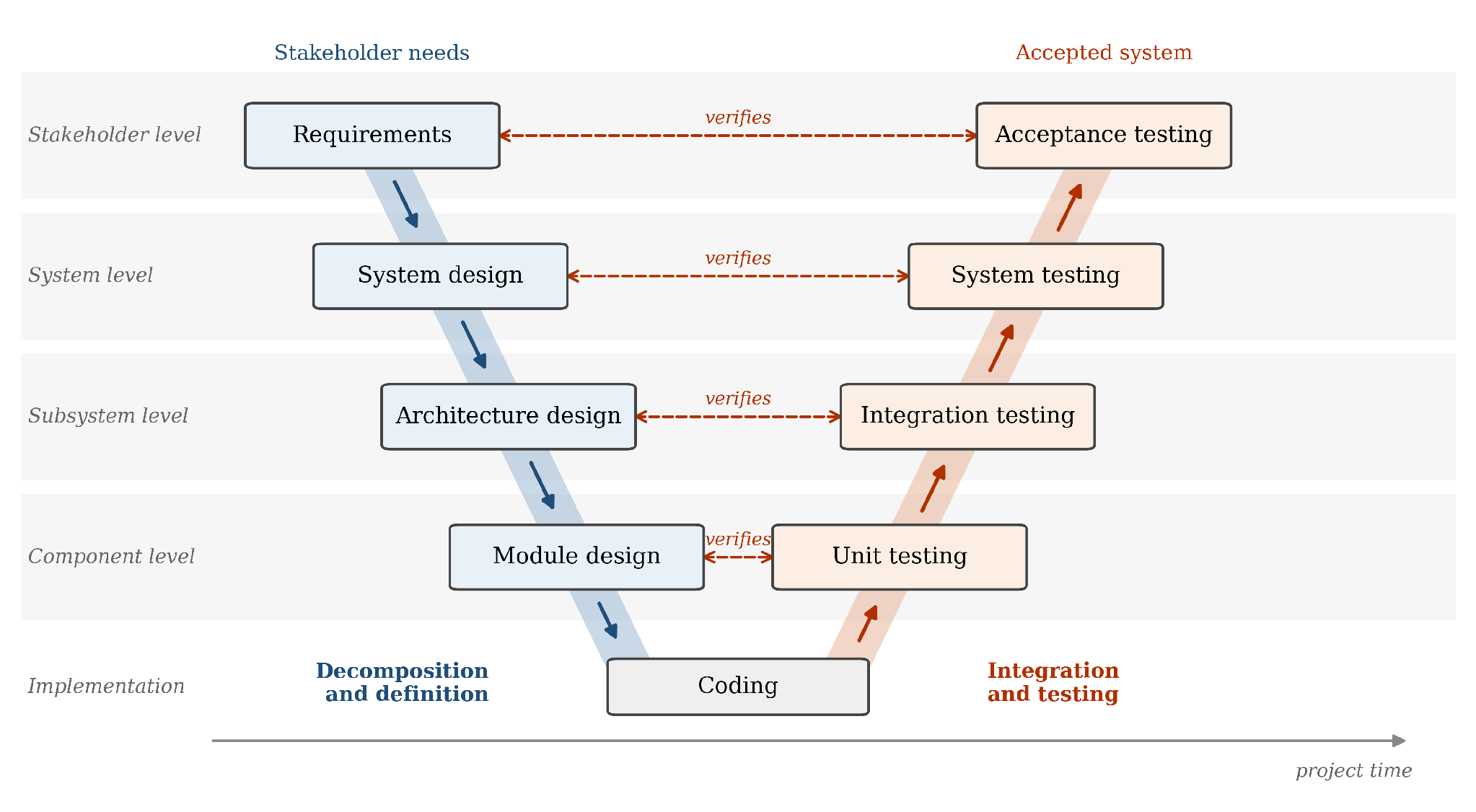}
\caption{The V-Model. The left arm decomposes stakeholder needs into progressively more detailed designs, and the right arm integrates and assesses the result in reverse order. Each dashed link pairs a design activity with the test activity that verifies it at the same level of abstraction}
\end{figure}

We use the V-Model's pairing principle to classify verification evidence. At each level, we ask whether the artifacts available for review contain evidence addressing the corresponding work, such as acceptance evidence for a requirement or unit-test evidence for modified logic (Boehm 1984). Verification may be recorded elsewhere, so a missing check in a pull request does not necessarily mean that no check occurred. VeeReview can assess only the pull-request artifacts and repository context it retrieves. Sections 5.1 and 6.5 evaluate whether these artifacts can serve as a proxy for verification practice.

\textbf{Unit of measurement.} Most pull requests do not change every level of abstraction. A single V-Model summary would therefore require decisions about how to weight unrelated activities and judge their applicability. We classify each build and verify pair separately. An applicable pair can be matched, build-only or verify-only. The three model-judged pairs can also be unverified when the evidence check cannot substantiate a claimed half. A fifth state, not applicable, marks pairs outside the change's scope. Table 1 defines these states. The preregistered outcome is the build-only rate among pairs classified as applicable. Because the instrument also decides applicability, the assessed subsets may differ between cohorts.

\begin{table}[!htbp]
\caption{Match states of a build and verify pair. Build-only is the state the study treats as leaving the pair open. The unverified state arises only on the three model-judged pairs (Section 3.4)}
\label{tab:1}
\footnotesize
\setlength{\tabcolsep}{4pt}
\begin{tabularx}{\textwidth}{lll>{\raggedright\arraybackslash}X}
\toprule
\textbf{Match state} & \textbf{Build half} & \textbf{Verify half} & \textbf{Meaning} \\
\midrule
matched & present & present & Evidence of the work and of the corresponding check is present \\
build-only & present & absent & Evidence of the work is present but evidence of the corresponding check is not \\
verify-only & absent & present & Evidence of a check is present without evidence of corresponding work at this level \\
unverified & claimed & claimed & The model claimed half was present but cited no evidence found in the pull request \\
not applicable & -- & -- & Neither half is relevant to the change \\
\bottomrule
\end{tabularx}
\end{table}

\textbf{The four pairs.} A pull-request diff and its context can provide evidence for four V-Model pairs. Table 2 defines each pair and its decision procedure, and Figure 2 illustrates them. They are ordered by abstraction level, rather than importance. A deterministic rule decides the fourth pair, Module Design $\leftrightarrow$ Unit Test, from the diff because changes to module logic and the presence of a test can be extracted directly. Its verdict is therefore unaffected by the wording of the pull request, a property used in Section 5.2.

\begin{table}[!htbp]
\caption{The four build and verify pairs scored by the instrument. The fourth pair is decided by a deterministic rule over the diff rather than by the model.}
\label{tab:2}
\footnotesize
\setlength{\tabcolsep}{4pt}
\begin{tabularx}{\textwidth}{ll>{\raggedright\arraybackslash}Xl}
\toprule
\textbf{Build half} & \textbf{Verify half} & \textbf{The question the pair asks} & \textbf{Decided by} \\
\midrule
Requirement & Acceptance & Is a requirement stated, and is there evidence it was accepted? & model, constrained by signals \\
System Design & System Test & Did system level behavior change, and is it exercised end to end? & model, constrained by signals \\
Architecture & Integration & Did a boundary or interface change, and is the integration exercised? & model, constrained by signals \\
Module Design & Unit Test & Did module logic change, and did the unit test arrive with it? & deterministic rule \\
\bottomrule
\end{tabularx}
\end{table}

\begin{figure}[!htbp]\centering
\includegraphics[width=\linewidth]{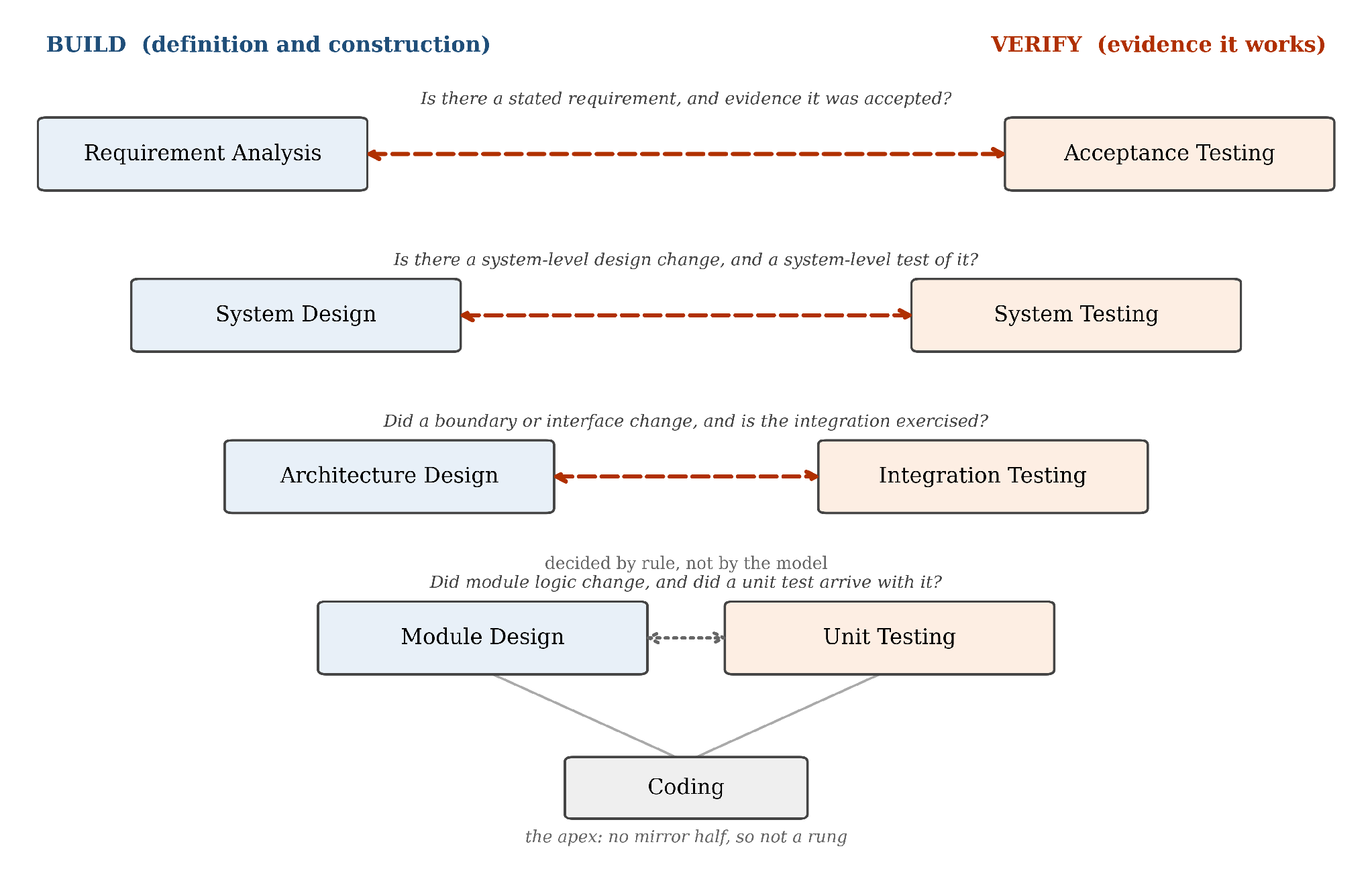}
\caption{The V-Model reduced to the claim this paper tests. Each pair joins a build activity to the verification activity at the same level. The fourth pair is decided by a deterministic rule over the diff. Coding, at the apex, has no mirror half and is therefore not a pair}
\end{figure}

\textbf{An illustrative example.} This example is constructed from the rules in Appendix A rather than taken from the corpus. Consider a bugfix pull request that changes one function, references the issue it addresses and adds no test. The issue reference supplies requirement evidence, but the acceptance half needs separate evidence that the reported problem was addressed. For Module Design $\leftrightarrow$ Unit Test, the rule detects modified production logic and no newly added test. If the repository tree contains a file meeting the heuristic for an existing related test, the pair is \textit{matched} and flagged as not updated. Otherwise, it is \textit{build-only}. A filename match does not establish that the existing test exercises the changed logic. If the change affects neither deployment, system-level behavior nor an interface, the two middle pairs are not applicable.

\textbf{Coding is not a pair.} Writing the code sits at the apex of the V, where development turns toward verification. It has no corresponding verification activity in the diagram, so treating it as a fifth pair would require inventing a partner.

\textbf{Scope of the claim.} We use build and verify relationships to organize evidence visible during pull-request review, rather than to measure adherence to the V-Model lifecycle. An existing test may satisfy the instrument's rule even if it was not created or updated with the change. Verification outside the retrieved artifacts may go undetected. Section 6.5 examines whether these distinctions were informative in this corpus.

\subsection{Agent-authored pull requests and the AIDev dataset}

AIDev provides the sampling frame for agent-authored changes (Li et al. 2026). It records 932,791 agent-authored pull requests across 116,211 repositories from OpenAI Codex, GitHub Copilot, Devin, Cursor and Claude Code. We sample its curated subset of 33,596 pull requests from 2,807 repositories (Section 4.2). Because AIDev does not provide a comparable human cohort, we construct one from GitHub (Section 4.3). For the selected pull requests, we retrieve diffs, reviews, CI results and linked-issue context (Section 3.2).

\textbf{Sampling rather than scoring the entire corpus.} Three properties shaped this decision. First, agent frequencies are uneven: a proportional 10\% sample would include only 46 Claude Code pull requests, limiting agent-specific analyses (Section 4.2). Second, agents cluster by repository. Only 62 of 958 repository clusters in the scored corpus contain more than one agent, and one repository accounts for 24.8\% of the corpus. More pull requests from the same clusters do not provide the same information as independent repository clusters. Third, the comparison requires eligible human-authored pull requests in the same repositories: 2,194 of 3,760 sampled agent pull requests had no eligible candidate (Section 4.3). Retrieval and scoring also have costs. At approximately 264 pull requests per hour (Appendix C), scoring the curated subset of 33,596 pull requests would take about 127 hours of local inference. These constraints explain our sampling design.

We adopt the "AI teammate" framing of Li et al. (2025). An agent that takes an issue, produces a change, opens a pull request and responds to review participates in the development process, rather than only assisting an author in an editor. This framing has two consequences for measurement.

First, the pull request is the unit of analysis. Its artifacts can provide evidence for both the build and verification halves. VeeReview uses material available through GitHub and does not require instrumentation of the agent.

Second, authorship workflows may produce different descriptions, issue references and other review artifacts. Agent harnesses may generate structured descriptions or issue links, while human contributors follow other conventions. We measure some of these observable differences, but do not observe harness settings or author intent directly. A model reading the pull request may therefore respond to documentation and available evidence as well as to verification work. Section 5.2 examines these associations. The rule-decided pair provides a comparison unaffected by description wording.

\section{VeeReview and Its Build and Verify Classification Procedure}

This section describes VeeReview's decision procedures and their inputs. Figure 3 summarizes the pipeline, distinguishing mechanically extracted signals, model-judged states and the rule-decided pair. Appendix C lists the implementation files, and Appendix A reproduces the judging rubric. These details identify which measurements depend on model judgement.

\begin{figure}[!htbp]\centering
\includegraphics[width=\linewidth]{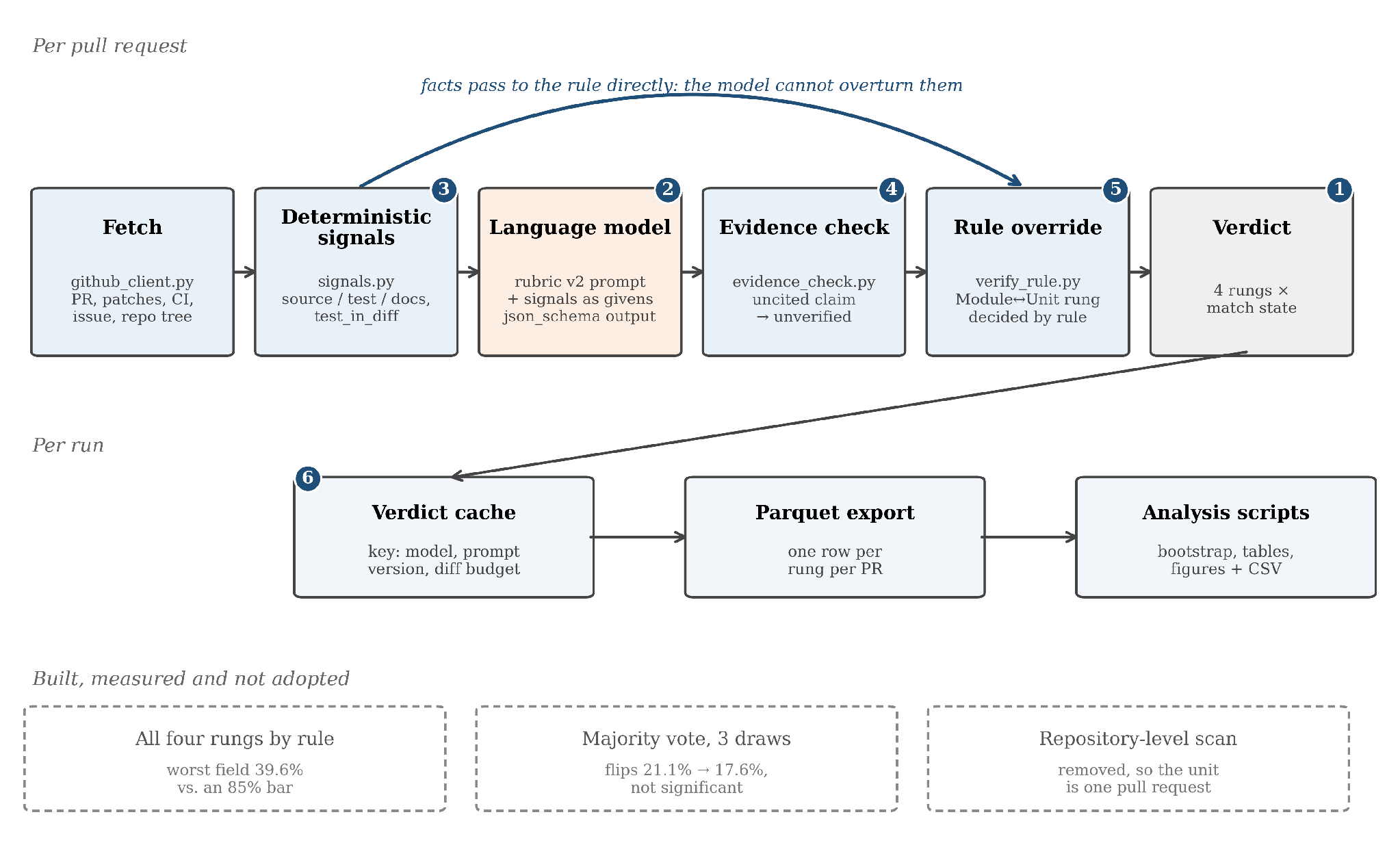}
\caption{The architecture of VeeReview. The top row is the path one pull request takes, in order of execution. Deterministic signals reach the rule override directly, so the model cannot overturn a fact established by rule. The bottom row is the run infrastructure that keeps a corpus of verdicts comparable}
\end{figure}

\subsection{Overview and definitions}

For a pull request \textit{p} and pair \textit{r}, B (\textit{p}, \textit{r}) denotes the presence of the build half and V (\textit{p}, \textit{r}) the presence of the verification half. The state is \textit{matched} when both hold, \textit{build-only} when only B holds, \textit{verify-only} when only V holds, and \textit{not applicable} (na in tables) when neither half is relevant to the change. Two refinements apply. For the three model-judged pairs, a claim that B or V holds must cite evidence present in the pull request. Otherwise, the state is \textit{unverified}. For Module Design $\leftrightarrow$ Unit Test, the model decides neither half. B holds when production logic changes, and the rule computes V from a test in the diff or an existing test identified as covering the changed paths (Section 3.5, Appendix A).

The instrument applies three layers. Deterministic signals constrain the model, and the rule for Module Design $\leftrightarrow$ Unit Test has the final say on that pair.

1. Deterministic signals are computed first and are never revised (Section 3.3).

2. The model accepts these signals and answers questions that require interpreting the diff and description. The instrument checks its cited evidence (Section 3.4).

3. A rule decides the Module Design $\leftrightarrow$ Unit Test pair, and the model may only explain that verdict (Section 3.5).

\subsection{Inputs}

\textbf{Inputs.} A diff alone cannot establish whether a change has appropriate verification evidence. VeeReview retrieves pull-request metadata, changed files and patches, review submissions, CI checks on the head commit and the linked-issue body when it can parse an issue reference. It also retrieves the repository tree at the head commit to find candidate pre-existing tests. This tree check identifies potential correspondence between files, rather than demonstrated coverage. For Module Design $\leftrightarrow$ Unit Test, the instrument distinguishes a newly added test, a heuristically identified existing test and no identified test (Section 3.5).

The model receives patches under a fixed budget of 120,000 characters for every reported run. The budget forms part of the cache key, so changing it requires rescoring every pull request rather than mixing input conditions. The export records whether each diff was truncated. Truncation is the main threat to RQ3 and is analyzed in Section 5.4.

\subsection{Deterministic signals}

Before calling the model, the signal extractor classifies changed paths as source, test, documentation or configuration. It also classifies tests by V-Model level (unit, integration or system), checks whether the diff contains a test, retrieves the CI conclusion on the head commit and parses linked issue numbers from the title, body and commits. Appendix B lists the heuristics.

Two signals are especially important. The \textit{test-in-diff} signal supplies the central evidence for the fourth pair. The \textit{comment-only} signal prevents a documentation edit inside a \texttt{.ts} or \texttt{.py} file from being counted as a source change. Without this exclusion, documentation-heavy pull requests can appear to contain development work that did not occur. The classifier parses patch hunks and excludes files whose changed lines consist only of comments or docstrings.

\subsection{Model-judged pairs and the evidence check}

For the three model-judged pairs, the judge receives deterministic signals and a rubric defining build and verification evidence. It returns a response under a constrained schema. The instrument checks whether the cited paths and artifacts appear in the retrieved material. This check validates references but does not assess whether a cited test exercises the changed behavior or whether acceptance evidence is adequate. Appendix A reproduces the rubric. Before the pair enumeration was constrained, a smaller model returned seven pairs with invented identifiers. The failure appeared downstream as "not applicable everywhere" rather than as a malformed response.

The evidence check applies to every model claim that a half is present. If the cited files or paths do not appear in the change, the instrument assigns \textit{unverified}. We report this as a separate state throughout the paper: an unsupported model assertion does not establish that verification is absent. Section 5.2 shows that the cohort difference in unverified classifications exceeds some of the pair-level differences.

\subsection{The rule-decided pair}

A deterministic rule assigns Module Design $\leftrightarrow$ Unit Test using changed production files, test files in the diff and filename-based evidence of a potentially related test in the repository tree. It replaces the model's verdict, although the model still supplies explanatory text. Appendix A gives the truth table. The match state is independent of model judgement and description wording but still depends on classification heuristics. Finding a potentially related test file does not establish coverage of the modified logic.

This separation permits two comparisons. We can compare the rule-decided state's repeatability with that of the model-judged states (Section 5.1) and repeat the cohort comparison on a measure unaffected by the judge's reading of prose (Section 5.2). Neither comparison validates the rule, whose accuracy depends on its test-identification heuristics.

\subsection{Outputs}

For each pair, VeeReview exports applicability, both halves' states, a match state, a score from 1 to 10, a color band, a short finding and supporting references. Matched, build-only and verify-only describe the evidence classified for a change, and not applicable marks a pair outside its scope. Unverified instead records a failure to substantiate a model claim. We retain and analyze it separately. Inferential analyses use match states. We summarize numerical scores only at cohort level because individual scores are unstable (Section 5.1).

\subsection{Design evolution}

We made six design changes in response to measured problems before reaching the version shown in Figure 3. We also built and evaluated three alternatives that we did not adopt. The replication package includes a methodology document recording the measurements behind these decisions. The adopted changes were:
\begin{itemize}
\item \textit{Four pairs instead of three.} The first version combined design and integration in one pair. We split it into System Design $\leftrightarrow$ System Test and Architecture $\leftrightarrow$ Integration to represent each abstraction level separately. Earlier labelled examples could not evaluate the two new pairs, motivating the new reference set in Section 4.5.
\item \textit{Grammar-constrained output.} For the same five pull requests, constrained decoding produced five schema-conformant results, while plain JSON mode produced none. The additional run time was about 6\%.
\item \textit{The comment-only classifier} (Section 3.3).
\item \textit{Evidence checks} (Section 3.4).
\item \textit{One rule-decided pair.} This pair changed state on 0\% of reruns, compared with 26\% to 39\% for the model-judged pairs (Section 5.1).
\item \textit{Configuration-aware caching. Including the model, prompt version and diff budget in the cache key} prevents verdicts from one configuration being reused under another.
\end{itemize}

We evaluated and dropped three alternatives. An all-rule classifier failed its preset agreement threshold: across 2,405 cached verdicts and a preset threshold of 85\%, the worst field reached 39.6\% agreement (the System Design $\leftrightarrow$ System Test match state). Mechanical verification halves reached 69\% to 86\% agreement because detecting a system-level test in a diff is relatively factual. Mechanical build halves, however, equated a configuration-file change with a system-level design decision. We did not tune the classifier to agree more closely with the model, since model agreement does not establish validity. The rules remain in the repository but are not connected to the pipeline. Majority voting over three model draws reduced match-state changes from 21.1\% to 17.6\%. We could not distinguish that reduction from zero at the evaluated sample size, so it did not justify about 37 hours of full-corpus inference. Finally, we removed a repository-level structural scan and a planned question about repository health to keep the study focused on individual pull requests.

\section{Study Design}

We preregistered the research questions, analysis choices and falsification criteria before the main analysis. The final study also includes documented deviations and analyses added after examining the original results. Sections 4.6 and 4.7 distinguish preregistered tests, robustness checks and exploratory analyses. Figure 4 shows the sampling, matching and scoring pipeline.

\begin{figure}[!htbp]\centering
\includegraphics[width=\linewidth]{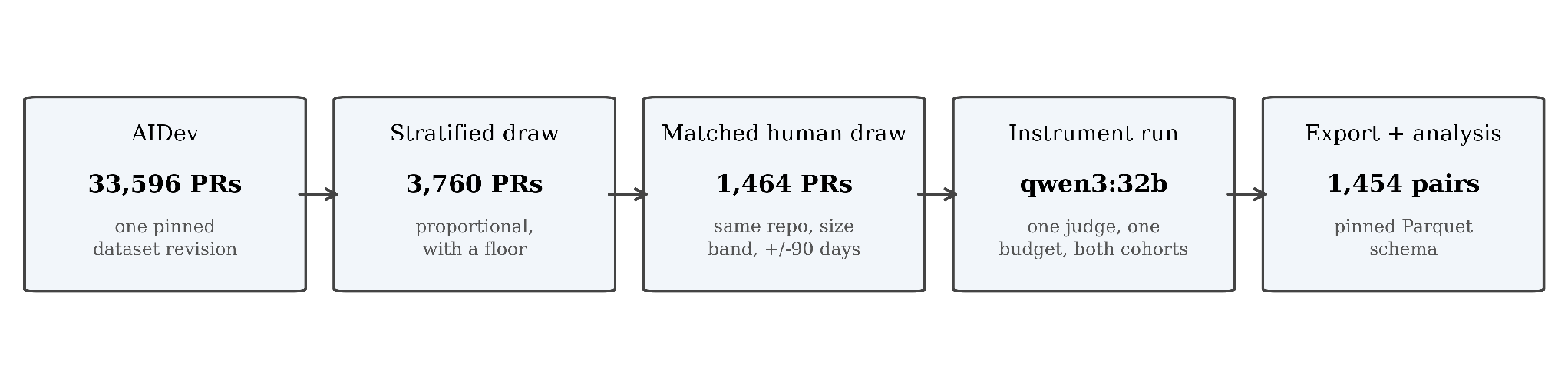}
\caption{From corpus to verdicts. Agent-authored pull requests are drawn from the AIDev curated subset, matched to human-authored pull requests within the same repository, and both cohorts are scored by one judge model under one configuration. Each stage is caught and resumable, and the sample and the dataset revision are frozen}
\end{figure}

\subsection{Research questions and hypotheses}

\textbf{RQ1 (Prevalence).} How do agent-authored and human-authored pull requests differ in each pair's match state? We preregistered the hypothesis that agent-authored pull requests would have higher build-only rates, especially at Module Design $\leftrightarrow$ Unit Test. The falsification rule rejected the hypothesis only if no pair showed a cohort difference. We also committed to reporting a reversed result if one occurred.

\textbf{RQ2 (Locus).} Which pair do agents most often leave build-only, and does this vary by pull request type and authoring agent? Equal build-only rates across pairs would indicate a general effect without a specific locus.

\textbf{RQ3 (Size).} How does pull request size relate to matching, and does the relationship differ by agent? We measure size in three ways: lines changed (the prespecified primary metric), diff characters (the model's input volume) and files touched (the change's breadth). These measures can diverge in practice.

\subsection{Agent cohort}

The agent cohort comes from a pinned revision of the AIDev curated subset: 33,596 pull requests across 2,807 repositories (Li et al. 2026). We draw 3,760 pull requests from 977 repositories, stratified proportionally by agent with a minimum stratum size. A proportional 10\% sample would leave Claude Code at n = 46, below RQ2's interpretability threshold, so we set the two smallest strata to 300. Each row carries a sampling weight: 10.00 for the three large strata, 5.14 for Cursor and 1.53 for Claude Code. We report aggregates pooled across agents both with and without weights. The sample is committed to the repository and reproduces byte for byte from its seed. Section 2.2 explains why we sample rather than score the full corpus.

\subsection{Matched human cohort}

AIDev has no matched human cohort. We therefore retrieved human-authored candidates from the same GitHub repositories within $\pm$90 days, giving priority to candidates in the same fixed diff-size band. Both cohorts were scored with the same judge model, prompt and character budget. This holds the scoring configuration constant and partially aligns repository context, time and size. It does not balance issue linking, test shipping, change purpose or unobserved verification practices. We fixed the following matching decisions before drawing the sample.
\begin{itemize}
\item \textit{One human pull request per agent pull request, without reuse.} Reusing a candidate would introduce dependence while making the comparison appear to have a larger sample.
\item \textit{Fixed size bands. Quartiles of the agent distribution} would make the matching criterion depend on the cohort being matched. We instead use the same fixed edges for matching and RQ3 reporting.
\item \textit{No match without an eligible human candidate. We drop the agent pull request }rather than create uneven matches and compensate with weights. This reduces coverage: of 3,760 sampled agent pull requests, 2,194 had no eligible candidate and 102 could not be sized. The remaining 1,464 matches yielded 1,454 complete scored pairs across 741 repositories, or 39\% of the agent sample.
\item \textit{Automation exclusion in three layers.} We exclude three groups: (i) bot accounts identified by API type, login suffix or a denylist, (ii) pull requests listed anywhere in AIDev, including those outside our sample, and (iii) contributors who drove an agent in the same repository. The third layer is necessary because AIDev records can associate an agent name with a human GitHub author. Filtering only the account type would therefore leave some agent-assisted work in the human cohort. This strict repository-specific exclusion reduces recall, especially in single-maintainer projects where the maintainer also drives the agent.
\end{itemize}

\textbf{The selection rule.} A human candidate must come from the same repository, fall within $\pm$90 days of the agent pull request and pass the automation exclusions. Assignment is greedy, with one human pull request per agent pull request and no reuse. Candidates are ranked by the triple (not in the same fixed size band, log-size distance |log(1+s) $-$ log(1+t) | when the bands differ and zero when they do not, absolute difference in creation time). We select the minimum triple. A candidate in the same band always takes priority. Time proximity breaks ties within a band, while log-size distance ranks candidates across bands.

The fallback uses log size because size differences are multiplicative: a 60-line candidate for a 50-line pull request is closer than a 5,000-line candidate for a 4,000-line pull request, although each lies one band away. Exact ties go to the first ranked candidate, deterministically given the frozen sample.

We process targets from the densest repositories first because an early assignment can consume a candidate needed by another target in the same repository. Processing in frozen sample order instead loses about 35 of 1,000 pairs, specifically in the repositories the design targets. Of the 1,464 matches, 814 (55.6\%) are in the agent pull request's size band and 650 use the log-size fallback. Section 5.2 reports the two strata separately.

\begin{table}[!htbp]
\caption{Composition of the matched sample. The two cohorts are matched one to one within repository on size band and a window of $\pm$90 days, so the repository count is shared by construction. The artifact rows and the not-applicable rates are what matching does not equalize.}
\label{tab:3}
\small
\setlength{\tabcolsep}{4pt}
\begin{tabular}{lll}
\toprule
 & \textbf{Agent cohort} & \textbf{Human cohort} \\
\midrule
Matched pairs & 1,454 & 1,454 \\
Repositories & 741 & 741 \\
Median lines changed & 84 & 65 \\
Mean lines changed & 1,259 & 466 \\
Links an issue & 29.7\% & 11.6\% \\
Ships tests & 32.5\% & 23.7\% \\
`na' rate, Requirement $\leftrightarrow$ Acceptance & 19.4\% & 28.1\% \\
`na' rate, System Design $\leftrightarrow$ System Test & 34.4\% & 32.9\% \\
`na' rate, Architecture $\leftrightarrow$ Integration & 46.1\% & 49.9\% \\
`na' rate, Module Design $\leftrightarrow$ Unit Test & 26.1\% & 20.6\% \\
\bottomrule
\end{tabular}
\end{table}

\subsection{Instrument configuration and run}

Both cohorts were scored on the same hardware by a locally served judge model (qwen3:32b), using prompt version v2, a 120,000-character diff budget, temperature zero and one draw per pull request. This holds the scoring configuration constant. We used one draw to limit cost, and Section 5.1 measures the resulting instability of individual verdicts. Appendix C lists the configuration. The authors used Claude and Claude Code (Anthropic) to help implement the instrument, write analysis and figure scripts, and edit the manuscript. They reviewed all AI-assisted output. The models did not determine the study design, analysis plan or interpretation. This use is also reported in the declarations.

\subsection{Instrument evaluation protocol}

We evaluated two properties: whether identical inputs yield the same states on rerun, and whether those states agree with independently assigned human labels. Configuration control and caching make scoring conditions comparable, but neither guarantees repeatability nor validates the labels. The two evaluations answer different questions.

\textbf{Consistency protocol.} We scored 19 pull requests three times each through the same endpoint, with an identical prompt at temperature zero. We checked agreement across draws for the overall score, the complete verdict (type and every pair's state, score and color) and each pair's match state.

\textbf{Calibration criterion.} Before running the study, we specified the following criterion for the reported match states:

\begin{quote}PASS when every applicable pair clears $\ge$ 80\% match state agreement against an independently human-labelled reference set, and there is no leniency drift toward \textit{matched} (that is, false matched $\le$ false unmatched), on a reference set of at least 20 and targeting 25--50 pull requests spanning types, labelled by two raters with an inter-rater statistic reported.\end{quote}

The leniency clause matters because a bias toward matched classifications could create a cohort difference if it affects the cohorts unequally. Symmetric error adds noise but does not by itself imply equal error for differently presented pull requests. We report Cohen's $\kappa$ (Cohen 1960) for inter-rater agreement and use 0.61, the conventional threshold for substantial agreement (Landis and Koch 1977), as the minimum for a reference-labelling procedure. Following software engineering reporting guidance (Díaz et al. 2023), we describe the unit of analysis, coding scheme and disagreement-resolution procedure alongside the statistic.

\textbf{Reference set.} Two raters independently labelled 25 pull requests: 13 agent-authored and 12 human-authored, randomly drawn and stratified by pull request type. Each pair within a pull request is one unit, giving 100 labels per rater. Raters chose matched, build-only, verify-only or not applicable without seeing the instrument's verdict or being told the cohort. Authorship could nevertheless be apparent from the account or description. The brief instructed them not to use merge outcomes. Neither rater contributed to the instrument or this paper.

The raters and first author resolved every disagreement jointly. Because the first author built the instrument, this involvement could influence the adjudicated labels. Section 5.1 therefore also reports agreement with each rater's original labels. The set exceeds the minimum of 20 pull requests and meets the lower end of the target range of 25 to 50.

Before calibration, we documented two defects in the deterministic procedure that could \textit{make it} too lenient. First, it counted a documentation-only edit in a \texttt{.ts} file as a source change. The comment-only classifier addresses this defect (Section 3.3). Second, the existing-test matcher accepts overlapping filename stems without establishing coverage of the changed logic. This defect remains, applies to both cohorts and is discussed in Section 7.

\subsection{Analysis methods}

Three data properties guide the analysis.

\textbf{Clustering.} Four pair rows share each pull request and pull requests cluster within repositories. One repository contributes 24.8\% of the agent corpus and the top five contribute 37.8\%. A per-row Wilson interval (Wilson 1927) treats all 14,592 rows as independent and is 1.8 to 7.3 times too narrow, depending on the pair. We therefore use repository-level bootstrap intervals (Efron and Tibshirani 1993; Cameron et al. 2008), except where a model-specific interval is stated. The released sidecars retain Wilson intervals for comparison.

\textbf{Pairing.} For RQ1, we also report an exact sign test on discordant matched pairs, where exactly one side received a matched state. The bootstrap accounts for repository clustering but not pairing, and the exact test accounts for pairing but not clustering. We report both to address these separate dependencies.

\textbf{Ordering.} RQ2 asks about the ordering of pair-level rates. We report how often repository-bootstrap resamples reproduce the observed order. Each iteration resamples repositories once for all pairs, preserving their shared repository structure.

\textbf{Interaction tests.} RQ2 and RQ3 ask whether profiles differ by agent, beyond differences in overall level. We construct a table with agents as rows and pairs (RQ2) or size bands (RQ3) as columns. For matched rate p$_{ij}$ in cell (i, j), let $\ell_{ij}$ = log (p$_{ij}$ / (1 $-$ p$_{ij}$)). We fit an additive row-plus-column model on this log-odds scale and calculate r$_{ij}$ = $\ell_{ij}$ $-$ $\ell_i\cdot$ $-$ $\ell\cdot_j$ + $\ell\cdot\cdot$, where a dot denotes a mean over that index. The statistic max$_{ij}$ |r$_{ij}$| is the largest cell departure from the agent's overall level and the column's overall difficulty. We use log odds because base rates differ greatly across columns. Additivity on the percentage scale could mistake curvature for interaction. Rates use unweighted counts, and a level enters the table only if every column has at least ten rows.

Agent identity is closely tied to repository membership (Section 5.3), so we generate two null distributions that preserve repository structure. \textbf{The within-repository} test uses only repositories containing multiple agents. Each permutation independently shuffles agent labels among pull requests within each repository, holding repository membership, size and the multiset of agent labels fixed.

The \textbf{across-repository tes}t first assigns every pull request its repository's modal agent, then shuffles those repository-level labels across repositories. Each repository keeps its pull requests, size and size profile. Only agent identity changes. We calculate the observed statistic on the same relabeled data used for the null distribution. Relabeling changes the agent for 139 of 3,648 pull requests, all in the 62 multi-agent repositories.

Both tests use 2,000 permutations with seed 20260906 and p = (hits + 1) / (permutations + 1). The four pair rows of a pull request always move together. Fifteen of the 958 repositories lack a unique modal agent. These receive the first agent's name alphabetically, independently of row order. We added both tests after the original analysis and report them as robustness analyses.

\textbf{Standardization.} We calculate directly standardized RQ1 rates (Rothman et al. 2008) to examine differences under a common distribution of observable pull-request characteristics. Each cohort is assigned the pooled mix of (links an issue) $\times$ (ships tests) $\times$ (size band). Because test shipping overlaps conceptually with the outcome for Module Design $\leftrightarrow$ Unit Test, we also report every standardized result using (links an issue) $\times$ (size band), that is, with test shipping removed. These adjusted comparisons describe conditional associations, rather than a causal decomposition.

\textbf{Preregistered model.} We fit the specified mixed-effects logistic regression with a repository random intercept and report odds ratios with correction across pairs in Section 5.2. The outcome is one assessed pair row, cohort is the fixed effect, and repository supplies the random intercept. Estimation uses maximum likelihood with the Laplace approximation (lme4). Wald intervals and p values come from this fit, with p values adjusted across the four pairs using Benjamini-Hochberg (Benjamini and Hochberg 1995).

The specification accounts for repository clustering but not explicitly for one-to-one matching. We therefore also report a conditional logistic regression stratified by matched pair and a generalized estimating equation with exchangeable correlation and repository clusters. The former conditions on pairing, and the latter estimates population-averaged effects with cluster-robust standard errors. Section 5.2 includes two further robustness fits: a crossed matched-pair random intercept and variational Bayes estimation of the preregistered specification.

We identify analyses developed after examining the data as exploratory where they appear.

\subsection{Deviations from the preregistration}

We document five departures from the preregistration in Appendix D. The following two affect the interpretation of the results.

\textbf{D1, statistics.} The original analysis used the repository bootstrap and exact paired test rather than fitting the preregistered mixed-effects logistic regression. We now report the specified model as the primary analysis in Section 5.2, alongside the bootstrap and sign test, which address clustering and pairing respectively. We apply the preregistered Benjamini-Hochberg correction across the four pairs (Benjamini and Hochberg 1995). This changes one conclusion: Architecture $\leftrightarrow$ Integration is not significant for the build-only outcome after adjustment.

\textbf{D2, three pairs became four.} The instrument has used four pairs since rubric v2, but the preregistration and falsification rules were written for three. The rule based on whether a confidence interval includes the null also cannot distinguish an effect in the hypothesized direction from an effect in the opposite direction. RQ1 produced the latter. We record this as a defect in the rule, rather than uncertainty about the result.

\section{Results}

The results use one judge model (qwen3:32b, prompt v2, diff budget 120,000 characters). The agent corpus contains 3,648 successfully scored pull requests, or 14,592 pair rows. RQ1 uses 1,454 complete pairs of agent and human pull requests across 741 repositories. Unless stated otherwise, intervals are 95\% repository-bootstrap intervals. Section 5.1 evaluates the instrument, and Sections 5.2 to 5.4 address the research questions in their preregistered order. Because the instrument did not meet its calibration criterion, the results in Sections 5.2 to 5.4 are reported as observations made with this instrument rather than as established cohort differences. Section 6 discusses their interpretation.

\subsection{Instrument evaluation}

\subsubsection{Rerun consistency}

Individual pull-request verdicts changed across identical runs. For 19 pull requests scored three times each under the conditions in Section 4.5:
\begin{itemize}
\item The overall score agreed across all three draws for 5 of 19 pull requests (26\%). Scores differed by at least a full point for 10 of 19.
\item The complete verdict was identical across all three draws for 33\% of pull requests.
\item Match state changed on 56 of 228 pair comparisons between identical reruns (25\%).
\end{itemize}

We did not isolate the implementation-level cause. Language-model output at temperature zero can be non-deterministic in code generation (Ouyang et al. 2025). Our observations are consistent with documented sources of instability: batch composition can change floating-point reduction order, while greedy decoding removes sampling noise but not arithmetic noise. Testing this explanation would require serial inference with a batch size of one. We judged the throughput cost too high and therefore report the observed variation without attributing it to a specific cause.

These reruns show substantial instability in individual verdicts. Changes had no consistent direction in the 19-pull-request experiment, but this small sample cannot rule out systematic or cohort-specific error. If scoring noise were independent and mean-zero, it would affect a large corpus mean less than an individual score. We did not verify that assumption for the matched comparison, so robustness of cohort estimates to repeated inference remains unestablished. In a separate comparison, the rule-decided pair changed state on 0\% of reruns, versus 26\% to 39\% for model-judged pairs. This establishes repeatability of the rule, rather than validity.

\subsubsection{Agreement with human raters}

VeeReview failed the preregistered agreement criterion. Agreement with adjudicated labels was 4\% for Requirement $\leftrightarrow$ Acceptance, 32\% for System Design $\leftrightarrow$ System Test, 60\% for Architecture $\leftrightarrow$ Integration and 44\% for Module Design $\leftrightarrow$ Unit Test. All were below the required 80\%. Agreement with each rater's original labels also remained below the threshold (Table 4). Substantial disagreement between raters means these figures establish poor agreement with the reference, rather than reliably estimating instrument accuracy.

We treat each rater's original labels as the primary references because the first author participated in adjudication. The adjudicated labels are reported alongside them. The replication package includes confusion matrices for every reference and pair. Disagreements are unevenly distributed: the dominant off-diagonal cell is a human label of matched with an instrument state of build-only. This accounts for 16 of the 25 adjudicated Requirement $\leftrightarrow$ Acceptance labels and explains the tendency to withhold matched credit described below.

\begin{table}[!htbp]
\caption{Calibration against the 25-pull-request reference set. The first four rows give the instrument's agreement with each reference, per pair, as a percentage of 25 labels. The preregistered bar is 80\%. The remaining rows give the direction of the instrument's errors and how they fall across the two cohorts. R1 and R2 are the two raters' own labels. All differences are agent minus human, in percentage points. Cohort drift is the amount the instrument's errors add to the agent cohort's matched rate relative to the human cohort's. The rater-reported gap is the cohort difference in the reference labels themselves.}
\label{tab:4}
\small
\setlength{\tabcolsep}{4pt}
\begin{tabularx}{\textwidth}{>{\raggedright\arraybackslash}Xlll}
\toprule
 & \textbf{R1} & \textbf{R2} & \textbf{Adjudicated} \\
\midrule
Requirement $\leftrightarrow$ Acceptance (\%) & 8 & 28 & 4 \\
System Design $\leftrightarrow$ System Test (\%) & 8 & 44 & 32 \\
Architecture $\leftrightarrow$ Integration (\%) & 48 & 72 & 60 \\
Module Design $\leftrightarrow$ Unit Test (\%) & 32 & 52 & 44 \\
Wrongly withheld matched (labels) & 48 & 16 & 30 \\
Wrongly awarded matched (labels) & 12 & 9 & 11 \\
Cohort drift, agent $-$ human (points) & $-$1 & +11 & +20 \\
Rater-reported gap, agent $-$ human (points) & +2 & $-$10 & $-$19 \\
\bottomrule
\end{tabularx}
\end{table}

\textbf{The reference set also failed its agreement requirement.} Overall inter-rater agreement was $\kappa$ = 0.154, with a maximum of 0.184 for any pair, well below the 0.61 minimum in Section 4.5. Because the raters often disagree, disagreement with the instrument cannot be attributed solely to instrument error. This calibration shows that VeeReview does not agree with these raters. It does not establish its accuracy.

\textbf{Error direction depends on the reference labels.} Against adjudicated labels, VeeReview assigned a non-matched state to 30 labels judged matched and assigned matched to 11 judged non-matched. Both original rater references show the same numerical direction (48 versus 12, and 16 versus 9). Within these references, the instrument thus withholds matched credit more often than it awards it.

\textbf{Appendix F reports two further patterns.} Against adjudicated labels, the instrument withholds matched credit more often for human-authored than agent-authored pull requests (38\% versus 23\% of labels). In this sample, that asymmetry would add about 20 percentage points to the apparent agent advantage. The pattern reverses against one rater's original labels (Table 4), and the raters disagree about the cohort difference itself. The reference therefore cannot confirm or rule out this asymmetry.

The instrument never assigns matched to Requirement $\leftrightarrow$ Acceptance in the reference set. Its 4\% agreement reflects this complete absence of matched classifications. One rater's labels for this pair also track merge outcomes despite the brief. Neither pattern accounts for the overall calibration failure: no pair reaches the threshold against any reference, even if Requirement $\leftrightarrow$ Acceptance is excluded.

\subsubsection{Interpretation scope}

The instrument agrees poorly with the available human labels, and the raters do not agree well enough to provide a reliable reference. We cannot separate instrument error from construct ambiguity or inconsistent human labelling. Relative to these labels, the instrument more often withholds a match, but this conditional pattern does not establish the direction of error against ground truth.

These findings constrain the remaining results. We report rates, differences and intervals as instrument-generated observations. The rule-decided pair provides the design's only internal comparison independent of the model's match-state judgement. The results support neither assessments of individual pull requests, an established agent advantage in Section 5.2, nor benchmark scores for agents. Repeatability in Section 5.1.1 does not establish accuracy. We also do not use the calibration asymmetry to correct the cohort estimates in Section 5.2 because its direction changes with the reference. We release the verdicts, reference set, both raters' labels and adjudication so others can reproduce the calibration and develop a stronger reference.

\subsection{RQ1 Prevalence}

We compare 1,454 agent-authored pull requests with their human partners on all four pairs. The preregistered primary outcome is build-only among pairs classified as applicable. We also report matched rates and the full five-state distribution because applicability and unverified classifications differ between cohorts. Tables 5 and 6 use different denominators to describe these complementary outcomes. Tables 7 and 8 standardize matched-rate comparisons to common distributions of selected observable characteristics.

\textbf{Build-only rate.} The agent cohort has lower rates on all four pairs: 64.8\% versus 74.0\% for Requirement $\leftrightarrow$ Acceptance, 56.8\% versus 70.1\% for System Design $\leftrightarrow$ System Test, 61.2\% versus 64.5\% for Architecture $\leftrightarrow$ Integration, and 42.3\% versus 48.4\% for Module Design $\leftrightarrow$ Unit Test. This reverses the hypothesized direction on every pair. The largest gap is at System Design $\leftrightarrow$ System Test, rather than at Module Design $\leftrightarrow$ Unit Test as predicted.

\textbf{Interval estimates.} Matched rates are higher for agents by +16.3 to +18.0 percentage points on the model-judged pairs and +6.1 points on the rule-decided pair (Table 5). Every repository-bootstrap interval excludes zero. The exact sign test agrees among discordant pairs, the agent side was matched in 77\% to 79\% of cases on the model-judged pairs. Mean overall scores also differ (4.38 versus 3.47), with a paired difference of +0.91 [+0.74, +1.08] across 741 repository clusters. As Section 5.1.1 explains, this is a cohort mean rather than an assessment of individual pairs.

\textbf{Preregistered model.} The maximum-likelihood mixed-effects model agrees with the direction on every pair. With a repository random intercept and cohort as the only fixed effect, agent-to-human odds ratios for a build-only verdict are 0.62 [0.52, 0.76] for Requirement $\leftrightarrow$ Acceptance, 0.52 [0.42, 0.63] for System Design $\leftrightarrow$ System Test, 0.86 [0.69, 1.07] for Architecture $\leftrightarrow$ Integration, and 0.70 [0.57, 0.86] for Module Design $\leftrightarrow$ Unit Test. Three pairs have adjusted p below 0.001 after Benjamini-Hochberg correction. \textbf{Architecture $\leftrightarrow$ Integration is not significant even before correction} (p = 0.17). Unlike the earlier risk-difference analysis, this model does not support a build-only effect for that pair. For the matched outcome, odds ratios are 3.91 [2.99, 5.11], 3.48 [2.69, 4.51], 3.63 [2.64, 5.00] and 1.43 [1.17, 1.76], all with adjusted p below 0.01.

Three robustness fits agree with the primary model. Conditional logistic regression stratified by matched pair gives build-only odds ratios of 0.61, 0.48, 0.75 and 0.68. Generalized estimating equations with repository clusters give 0.65, 0.56, 0.87 and 0.79. Variational Bayes estimation gives 0.62, 0.51, 0.85 and 0.69. Adding a crossed matched-pair random intercept leaves estimates unchanged, with pair-level variance at or near zero. In these fits, the repository term accounts for the dependence introduced by the design. We retain the conditional model for readers who prefer to condition directly on pairing.

\begin{table}[!htbp]
\caption{RQ1: matched-rate difference (agent $-$ human) per pair. The interval is a 95\% bootstrap over repositories computed on the difference itself. "Discordant a/h" counts pairs in which exactly one side matched, and the p value is from an exact two-sided sign test conditioned on those pairs. \textsuperscript{a} Decided by rule, not by the model}
\label{tab:5}
\footnotesize
\setlength{\tabcolsep}{4pt}
\begin{tabularx}{\textwidth}{>{\hsize=1.700\hsize\raggedright\arraybackslash}X>{\hsize=0.860\hsize\centering\arraybackslash}X>{\hsize=0.860\hsize\centering\arraybackslash}X>{\hsize=0.860\hsize\centering\arraybackslash}X>{\hsize=0.860\hsize\centering\arraybackslash}X>{\hsize=0.860\hsize\centering\arraybackslash}X}
\toprule
\textbf{Pair} & \textbf{Gap (pp)} & \textbf{95\% CI} & \textbf{Discordant a/h} & \textbf{Agent share} & \textbf{p} \\
\midrule
Requirement $\leftrightarrow$ Acceptance & +16.8 & [+13.3, +20.2] & 208/55 & 79.1\% & <10$^{-9}$ \\
System Design $\leftrightarrow$ System Test & +18.0 & [+14.1, +22.1] & 162/44 & 78.6\% & <10$^{-9}$ \\
Architecture $\leftrightarrow$ Integration & +16.3 & [+12.4, +20.1] & 101/30 & 77.1\% & <10$^{-9}$ \\
Module Design $\leftrightarrow$ Unit Test a & +6.1 & [+2.4, +9.8] & 168/114 & 59.6\% & 0.0016 \\
\bottomrule
\end{tabularx}
\end{table}

\begin{figure}[!htbp]\centering
\includegraphics[width=\linewidth]{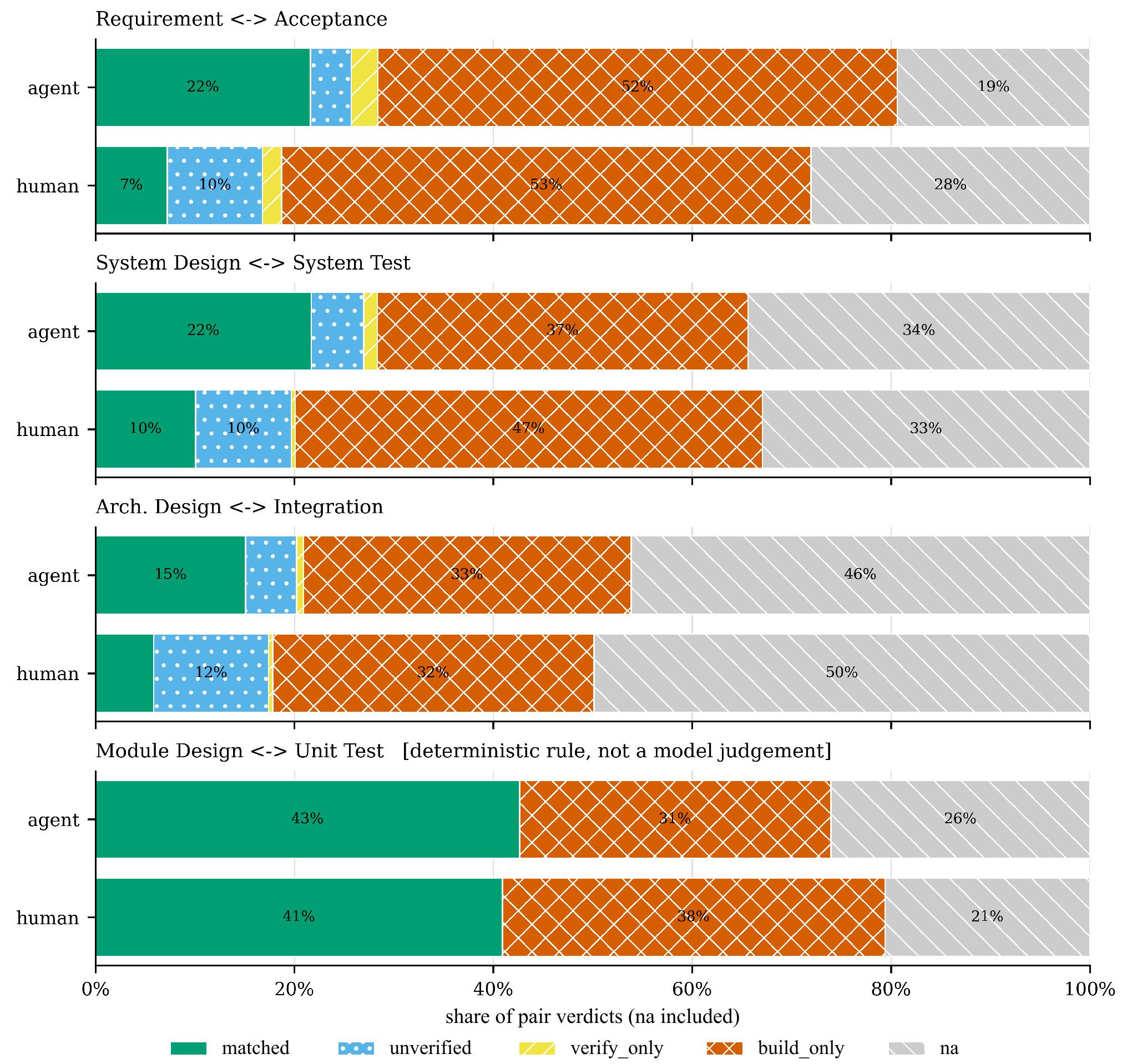}
\caption{RQ1: the full match state distribution per pair and cohort over all 1,454 pairs on each side, so that the not-applicable shares are comparable. Match state is encoded by hue, hatching and a fixed best-to-worst ordering, so the figure survives greyscale reproduction}
\end{figure}

\textbf{The cohorts have different assessed denominators.} We preregistered reporting the full state distribution because unequal not-applicable rates can confound conditional comparisons (Figure 5, Table 6). For Requirement $\leftrightarrow$ Acceptance, the judge assigns not applicable to 19.4\% of agent rows and 28.1\% of human rows, and the direction reverses for Module Design $\leftrightarrow$ Unit Test. Human pull requests also receive \textit{unverified}, meaning the model could not substantiate its cited evidence, about twice as often on each model-judged pair: 9.6\% versus 4.1\%, 9.6\% versus 5.3\%, and 11.6\% versus 5.2\%. Each reported conditional rate therefore uses its own assessed subset.

\begin{table}[!htbp]
\caption{RQ1: the full five-state distribution per pair and cohort, as a percentage of all 1,454 pairs. Percentages are over every pair, not-applicable included, because the cohorts do not draw not-applicable equally often.}
\label{tab:6}
\footnotesize
\setlength{\tabcolsep}{4pt}
\begin{tabularx}{\textwidth}{>{\hsize=1.700\hsize\raggedright\arraybackslash}X>{\hsize=0.883\hsize\centering\arraybackslash}X>{\hsize=0.883\hsize\centering\arraybackslash}X>{\hsize=0.883\hsize\centering\arraybackslash}X>{\hsize=0.883\hsize\centering\arraybackslash}X>{\hsize=0.883\hsize\centering\arraybackslash}X>{\hsize=0.883\hsize\centering\arraybackslash}X}
\toprule
\textbf{Pair} & \textbf{Cohort} & \textbf{matched} & \textbf{unverified} & \textbf{verify-only} & \textbf{build-only} & \textbf{na} \\
\midrule
Requirement $\leftrightarrow$ Acceptance & agent & 21.6 & 4.1 & 2.7 & 52.2 & 19.4 \\
 & human & 7.2 & 9.6 & 1.9 & 53.2 & 28.1 \\
System Design $\leftrightarrow$ System Test & agent & 21.7 & 5.3 & 1.4 & 37.3 & 34.4 \\
 & human & 10.0 & 9.6 & 0.4 & 47.0 & 32.9 \\
Architecture $\leftrightarrow$ Integration & agent & 15.1 & 5.2 & 0.7 & 32.9 & 46.1 \\
 & human & 5.8 & 11.6 & 0.4 & 32.3 & 49.9 \\
Module Design $\leftrightarrow$ Unit Test & agent & 42.6 & 0.0 & 0.0 & 31.3 & 26.1 \\
 & human & 40.9 & 0.0 & 0.0 & 38.4 & 20.6 \\
\bottomrule
\end{tabularx}
\end{table}

\textbf{Standardization on observable characteristics.} Agent pull requests link issues in 29.7\% of pairs versus 11.6\% for human partners, and ship tests in 32.5\% versus 23.7\% (Table 3). Standardizing matched rates to the pooled mix of (links an issue) $\times$ (ships tests) $\times$ (size band) leaves differences of +6.8 points [+3.9, +10.2] for Requirement $\leftrightarrow$ Acceptance, +10.7 [+7.2, +14.7] for System Design $\leftrightarrow$ System Test, and +10.8 [+8.2, +14.5] for Architecture $\leftrightarrow$ Integration. These retain 40\%, 59\% and 66\% of the crude differences. The rule-decided difference changes from +6.1 to $-$1.2 points [$-$3.9, +2.4].

Because test shipping overlaps with the outcome for Module Design $\leftrightarrow$ Unit Test, we repeat the analysis without it. The model-judged differences are then +7.2 [+4.2, +10.3], +12.7 [+8.7, +17.0] and +13.0 [+9.4, +17.0], retaining 43\%, 71\% and 79\% of the crude gaps. The rule-decided difference is +3.1 points [$-$0.4, +7.1]. All intervals are repository bootstraps of the standardized difference, and Table 8 reports both specifications. The rule-decided interval includes zero under either specification, while the model-judged intervals exclude it under both.

\begin{table}[!htbp]
\caption{RQ1: the matched-rate difference before and after direct standardization to the pooled mix of (links an issue) $\times$ (ships tests) $\times$ (size band). Rates and differences are in percentage points. \textsuperscript{a} Decided by rule, not by the model}
\label{tab:7}
\footnotesize
\setlength{\tabcolsep}{4pt}
\begin{tabularx}{\textwidth}{>{\hsize=1.700\hsize\raggedright\arraybackslash}X>{\hsize=0.860\hsize\centering\arraybackslash}X>{\hsize=0.860\hsize\centering\arraybackslash}X>{\hsize=0.860\hsize\centering\arraybackslash}X>{\hsize=0.860\hsize\centering\arraybackslash}X>{\hsize=0.860\hsize\centering\arraybackslash}X}
\toprule
\textbf{Pair} & \textbf{Agent} & \textbf{Human} & \textbf{Crude gap} & \textbf{Standardized gap} & \textbf{Retained} \\
\midrule
Requirement $\leftrightarrow$ Acceptance & 26.8 & 10.0 & +16.8 & +6.8 & 40\% \\
System Design $\leftrightarrow$ System Test & 33.0 & 15.0 & +18.0 & +10.7 & 59\% \\
Architecture $\leftrightarrow$ Integration & 28.0 & 11.7 & +16.3 & +10.8 & 66\% \\
Module Design $\leftrightarrow$ Unit Test a & 57.7 & 51.6 & +6.1 & -1.2 & --- \\
\bottomrule
\end{tabularx}
\end{table}

\begin{table}[!htbp]
\caption{RQ1: the matched-rate difference after direct standardization under two stratum definitions. Differences are in percentage points, with 95\% repository-level bootstrap intervals computed on the standardized difference. The second definition drops test shipping from the stratum because it is conceptually adjacent to the Module Design $\leftrightarrow$ Unit Test outcome. Retained is the standardized difference as a share of the crude one, under each definition in turn. \textsuperscript{a} Decided by rule, not by the model}
\label{tab:8}
\small
\setlength{\tabcolsep}{4pt}
\begin{tabularx}{\textwidth}{>{\hsize=1.700\hsize\raggedright\arraybackslash}X>{\hsize=0.825\hsize\centering\arraybackslash}X>{\hsize=0.825\hsize\centering\arraybackslash}X>{\hsize=0.825\hsize\centering\arraybackslash}X>{\hsize=0.825\hsize\centering\arraybackslash}X}
\toprule
\textbf{Pair} & \textbf{Crude} & \textbf{Standardized, issue $\times$ tests $\times$ size} & \textbf{Standardized, issue $\times$ size} & \textbf{Retained} \\
\midrule
Requirement $\leftrightarrow$ Acceptance & +16.8 & +6.8 [+3.9, +10.2] & +7.2 [+4.2, +10.3] & 40\% / 43\% \\
System Design $\leftrightarrow$ System Test & +18.0 & +10.7 [+7.2, +14.7] & +12.7 [+8.7, +17.0] & 59\% / 71\% \\
Architecture $\leftrightarrow$ Integration & +16.3 & +10.8 [+8.2, +14.5] & +13.0 [+9.4, +17.0] & 66\% / 79\% \\
Module Design $\leftrightarrow$ Unit Test \textsuperscript{a} & +6.1 & $-$1.2 [$-$3.9, +2.4] & +3.1 [$-$0.4, +7.1] & --- / 51\% \\
\bottomrule
\end{tabularx}
\end{table}

\textbf{Comparison with the rule-decided pair.} The mean crude difference across model-judged pairs is +17.0 points, versus +6.1 for the rule-decided pair, a ratio of 2.8. After standardization, the rule-decided difference is $-$1.2 points [$-$3.9, +2.4] with the full stratum and +3.1 points [$-$0.4, +7.1] without test shipping. Neither interval excludes zero. Module Design $\leftrightarrow$ Unit Test is decided from the diff and repository tree, so its state cannot change with description wording. It does not show the standardized difference seen on the three model-judged pairs. We specified this contrast during analysis and identify it as exploratory in Appendix D.

\textbf{Sensitivity analyses.} We precommitted two checks. For matching quality, 814 of 1,464 matches (55.6\%) fell in the agent pull request's size band, while 650 used the log-size fallback. Separate bootstraps show differences excluding zero for every pair in both strata, with overlapping intervals across strata (Figure 6). For size, differences appear in all five bands and are smallest in XS (under 50 lines) and largest in L (1,000 to 5,000 lines) on each model-judged pair (Figure 7). Their presence across bands argues against an explanation confined to one size band, although it does not remove all matching-related threats.

\textbf{Support for the standardization.} The full specification has 12 populated cells per pair. Of these, 11 contain at least five pull requests from each cohort and account for 98.9\% of the pooled weight. Both cohorts are represented in every cell used, meeting the method's positivity requirement. The reduced specification has six cells, all supported by both cohorts. The smallest cell contributing to a standardized rate contains five pull requests, the threshold applied by the analysis code.

\textbf{The pair count.} The preregistration used three pairs (Appendix D, D2), so we also reconstruct that rubric from the current outputs. This is not a rerun: rubric v1 used one design integration pair, and exact reproduction would require rescoring the corpus. Appendix D explains how we reconstruct v1 by recombining the two middle pairs' halves, which correspond to v1's question. Under the permissive mapping, either sub-pair can establish the combined build half, and either verification half can establish the combined verify half. The differences are +16.8 points [13.3, 20.2] for Requirement $\leftrightarrow$ Acceptance, +18.5 [14.9, 22.1] for the combined design-integration pair and +6.1 [2.4, 9.9] for Module Design $\leftrightarrow$ Unit Test, with every sign test below p = 0.002. Under the strict mapping, both sub-pairs must be matched. The combined difference is +10.2 points (15.4\% versus 5.3\%, with 144 discordant pairs favoring agents versus 37, p < 0.001). Both reconstructions preserve the direction and significance of the measured result.

\textbf{Selection into the matched sample.} The matched subset contains 1,454 of 3,648 scored agent pull requests and is not a random sample of the corpus. Compared with the 2,194 not retained, it links issues more often (29.7\% versus 5.6\%), ships tests less often (32.5\% versus 45.8\%) and is merged less often (55.8\% versus 79.4\%). Its repositories contribute 2.0 sampled agent pull requests each, compared with 7.7 for the rest. The agent mixes changes substantially: OpenAI Codex accounts for 29.5\% of the retained subset but 78.5\% of the rest, reflecting concentration in repositories without eligible human partners. RQ1 describes this selected subset, a limitation discussed in Section 7.

\textbf{Weighted and unweighted estimates.} Matching changes the agent mix, so we repeat RQ1 with sampling weights. Weighted differences are +18.7 points [15.0, 22.4], +17.8 [13.7, 21.9], +16.7 [12.6, 20.9] and +6.0 [2.1, 10.2], compared with unweighted differences of +16.8, +18.0, +16.3 and +6.1. The conclusions are unchanged. Both sets estimate RQ1 differences in the matched subset. We use weighting as a sensitivity check, rather than treating it as an estimate for the AIDev agent population.

\begin{figure}[!htbp]\centering
\includegraphics[width=\linewidth]{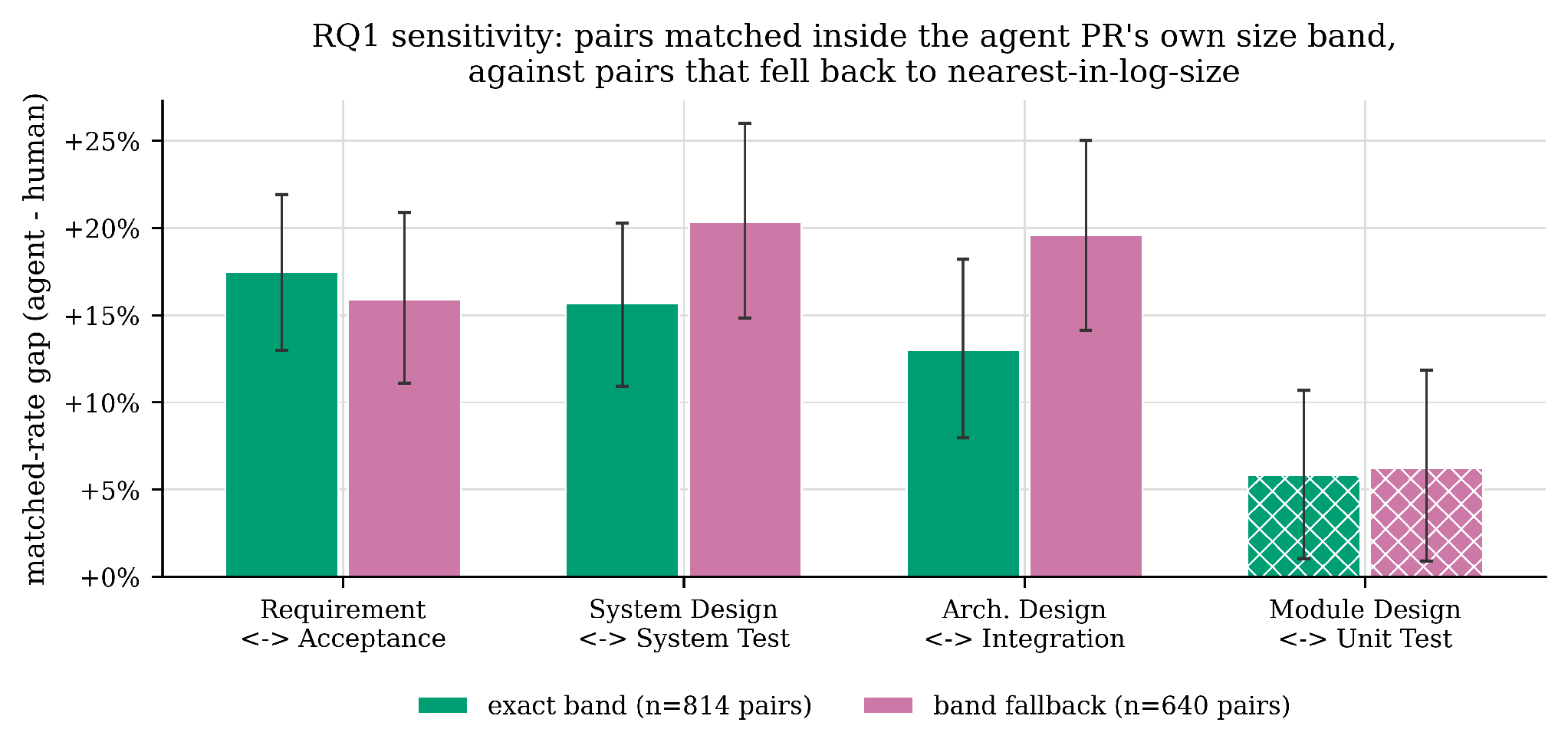}
\caption{RQ1 sensitivity to matching quality: the difference computed separately over pairs matched inside the agent pull request's own size band and over pairs that fell back to the nearest match in log size}
\end{figure}

\begin{figure}[!htbp]\centering
\includegraphics[width=\linewidth]{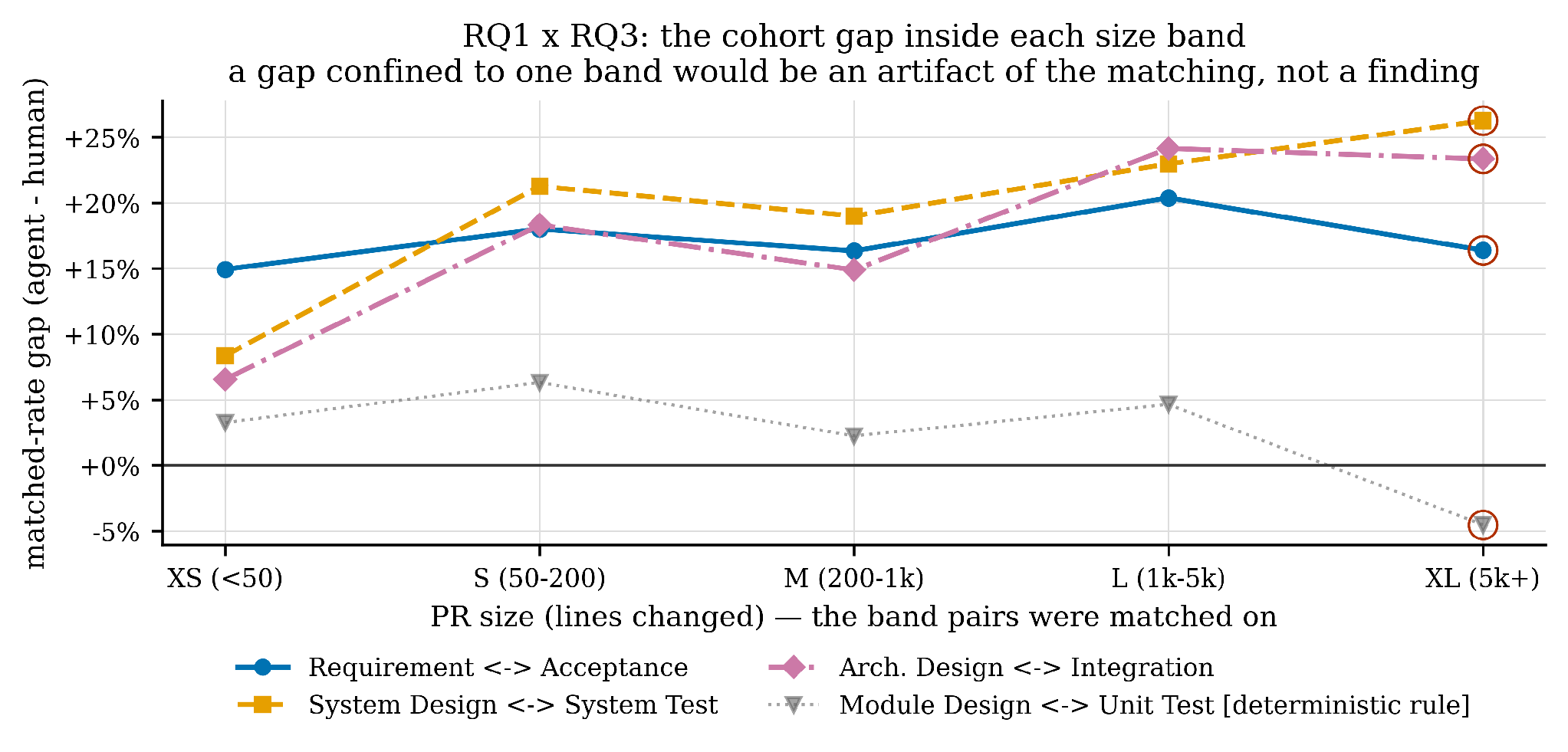}
\caption{RQ1 by size band: the cohort difference inside each size band. Circled points are cells below n = 30 on either side, which are reported but not interpreted}
\end{figure}

\textbf{Against the falsification rule.} The preregistered rule would reject the RQ1 hypothesis only if no pair showed a cohort effect. The four-point estimates instead run in the opposite direction. A rule based solely on whether an interval includes the null does not distinguish these outcomes (Section 4.7).

\textbf{Answer to RQ1.} VeeReview assigns lower build-only rates to agent pull requests on all four pairs. Preregistered odds ratios range from 0.52 to 0.86, with three pairs significant after correction, and all bootstrap intervals for matched-rate differences exclude zero. Standardization reduces the differences by 34\% to 60\%, and the rule-decided interval includes zero under both specifications. Both reconstructions of the earlier three-pair rubric preserve the direction and significance. This result is an observation made with this instrument on the matched subset, subject to the calibration result (Section 5.1) and the selection into matching (Section 7).

\subsection{RQ2 Locus}

For RQ2, we use all 3,648 scored agent pull requests. We calculate each pair's build-only rate among assessed rows, with repository-bootstrap intervals, and the proportion of resamples preserving the pair ordering. We then examine rates by authoring agent and pull request type. Table 9 reports rates and rank stability, and Figure 8 shows agent-specific profiles.

Requirement $\leftrightarrow$ Acceptance has the highest instrument-assigned build-only rate, at 71.6\% of applicable rows. Module Design $\leftrightarrow$ Unit Test has the lowest, at 37.2\% (Table 9). These are classification rates, rather than validated frequencies of missing verification.

\begin{table}[!htbp]
\caption{\textbf{Table 9 RQ2: build-only rate per pair over the agent corpus}. Rates are computed over assessed rows (not-applicable excluded), and the interval is a bootstrap over repositories. "Rank held" is the share of bootstrap resamples in which the pair keeps its observed rank. \textsuperscript{a} Decided by rule, not by the model}
\label{tab:9}
\footnotesize
\setlength{\tabcolsep}{4pt}
\begin{tabularx}{\textwidth}{>{\hsize=1.700\hsize\raggedright\arraybackslash}X>{\hsize=0.860\hsize\centering\arraybackslash}X>{\hsize=0.860\hsize\centering\arraybackslash}X>{\hsize=0.860\hsize\centering\arraybackslash}X>{\hsize=0.860\hsize\centering\arraybackslash}X>{\hsize=0.860\hsize\centering\arraybackslash}X}
\toprule
\textbf{Pair} & \textbf{build-only \%} & \textbf{95\% CI} & \textbf{n} & \textbf{Rank} & \textbf{Rank held} \\
\midrule
Requirement $\leftrightarrow$ Acceptance & 71.6 & [68.1, 74.3] & 2,872 & 1 & 100.0\% \\
System Design $\leftrightarrow$ System Test & 50.9 & [38.7, 64.5] & 2,468 & 3 & 78.9\% \\
Architecture $\leftrightarrow$ Integration & 57.4 & [53.0, 64.5] & 2,002 & 2 & 78.9\% \\
Module Design $\leftrightarrow$ Unit Test a & 37.2 & [27.0, 49.6] & 2,644 & 4 & 100.0\% \\
\bottomrule
\end{tabularx}
\end{table}

\textbf{Stability of the ordering.} Repository-bootstrap resamples preserve the full order in 78.9\% of iterations. Requirement $\leftrightarrow$ Acceptance remains highest and Module Design $\leftrightarrow$ Unit Test lowest in every resample. The middle pairs exchange places in approximately 21\% of resamples, so their order is less clear.

\textbf{Variation by authoring agent.} Pair profiles differ by agent under both tests that preserve repository structure. Only 62 of 958 repositories contain multiple agents, so shuffling labels across individual pull requests would create combinations of agents and repositories absent from the data. We use the null distributions defined in Section 4.6. Within the 62 multi-agent repositories (1,940 pair rows, 62 clusters), the observed interaction statistic is 1.11, above the null's 95th percentile of 0.90 (p = 0.005, 2,000 permutations). Across repositories, using each repository's modal agent (14,592 pair rows, 958 clusters), the statistic is 1.22 against 1.10 (p = 0.003, 2,000 permutations).

The within-repository test offers a cleaner comparison but relies on only 62 clusters, rather than the 958 underlying the corpus rates. The interaction remains detectable within well-powered pull request types (p = 0.001 for feature and p = 0.003 for bugfix). An exploratory restriction to pull requests without linked issues no longer detects it (p = 0.068). This does not show that issue links cause the difference. It suggests that visible pull-request artifacts may contribute to differences between agents, a possibility examined further in the discussion.

\begin{figure}[!htbp]\centering
\includegraphics[width=\linewidth]{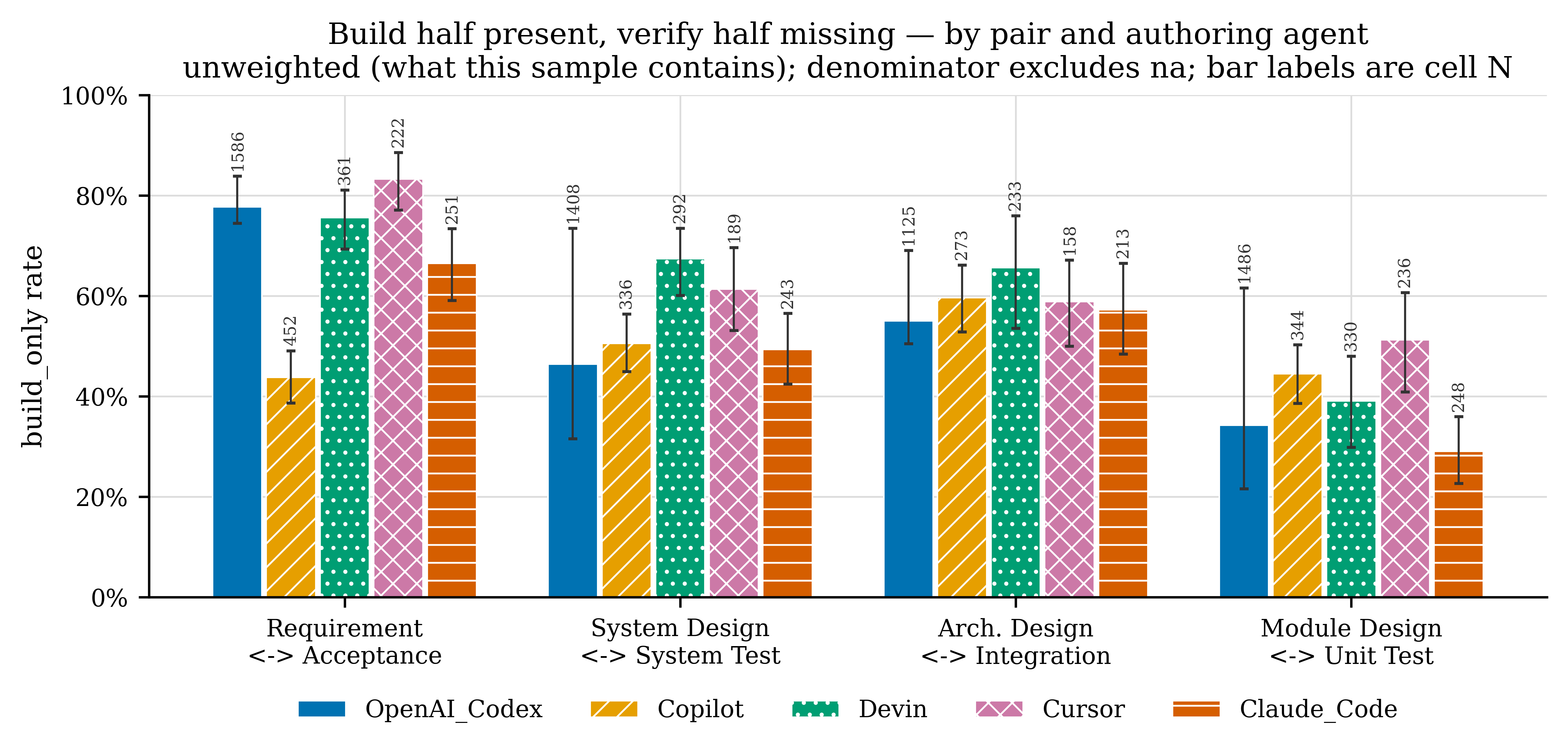}
\caption{\textbf{Fig. 8 RQ2: build-only rate per pair and authoring agent, unweighted. Each agent carries a hatch pattern as well as a color, so the figure survives greyscale reproduction.} Bar labels give cell n, and a starred label marks a cell below n = 30, which is reported but not interpreted}
\end{figure}

Observable profiles also differ. Codex links an issue in 1.0\% of pull requests, compared with 80.7\% for Copilot. Median size ranges from 65 lines for Codex to 445 for Claude Code, and test shipping from 22.0\% for Cursor to 45.8\% for Codex. Repository membership confounds these comparisons: of the 958 repositories, only 62 contain multiple agents, accounting for 485 pull requests and 13.3\% of pair rows. Agent comparisons therefore largely compare different repositories. We report them descriptively, rather than as rankings.

\textbf{Audit of the model's explanations.} We checked selected generated claims against extracted signals. Of 879 findings asserting that no requirement was stated, one (0.1\%) referred to a pull request with a parsed issue link. Of 147 findings mentioning CI, 56 (38.1\%) accompanied a passing CI status. This count does not establish contradiction because the audit did not distinguish assertions of CI success from assertions of failure. In contrast, 21 of 77 findings saying no test was added (27.3\%) accompanied a diff containing a test file, conflicting with the extracted signal. Explanations therefore need checks of what they assert, as well as which artifacts they cite. This audit does not validate the underlying pair states.

\textbf{Weighted sensitivity.} We repeat pooled rates with sampling weights to account for the stratum floors. Differences appear at the third decimal, and the ordering remains unchanged. Requirement $\leftrightarrow$ Acceptance stays at rank 1, with a weighted rate of 71.5\% versus 71.6\% unweighted. Module Design $\leftrightarrow$ Unit Test remains lowest at 37.3\% versus 37.2\%. Table 9 and Figure 8 report unweighted descriptive rates.

\textbf{Answer to RQ2.} Requirement $\leftrightarrow$ Acceptance has the highest instrument-assigned build-only rate (71.6\% of assessed rows), and Module Design $\leftrightarrow$ Unit Test the lowest (37.2\%). These endpoints are stable in the repository bootstrap, while the middle pairs cannot be reliably ordered. Profiles differ by agent, but strong associations between agents and repositories limit interpretation as agent-specific behavior. Failed calibration makes the pair ranking provisional.

\subsection{RQ3 Size}

For RQ3, we divide the agent corpus into five bands by lines changed. Within each band, we calculate match-state shares and the proportion of rows exceeding the diff budget. We repeat the analysis using diff characters and files touched. Table 10 reports the line-based results.

The matched share of all pair rows rises from 13.3\% below 50 lines to 38.0\% between 1,000 and 5,000 lines, then falls to 32.5\% above 5,000 lines. Not-applicable classifications decline from 45.8\% in the smallest band to 18.5\% in the 1,000-to-5,000-line band. Part of the higher matched share therefore reflects a change in which pairs are applicable, rather than necessarily better verification.

\begin{table}[!htbp]
\caption{\textbf{Table 10 RQ3: match state against pull request size} over the agent corpus, with the truncation rate per band. Truncation is caused by size and is therefore inseparable from the predictor, and the last column bounds the resulting artifact.}
\label{tab:10}
\footnotesize
\setlength{\tabcolsep}{4pt}
\begin{tabularx}{\textwidth}{>{\hsize=1.700\hsize\raggedright\arraybackslash}X>{\hsize=0.860\hsize\centering\arraybackslash}X>{\hsize=0.860\hsize\centering\arraybackslash}X>{\hsize=0.860\hsize\centering\arraybackslash}X>{\hsize=0.860\hsize\centering\arraybackslash}X>{\hsize=0.860\hsize\centering\arraybackslash}X}
\toprule
\textbf{PR size (lines changed)} & \textbf{matched \%} & \textbf{build-only \%} & \textbf{na \%} & \textbf{n rows} & \textbf{truncated} \\
\midrule
XS (<50) & 13.3 & 36.1 & 45.8 & 5,844 & 0.0\% \\
S (50-200) & 27.2 & 39.6 & 24.6 & 4,164 & 0.0\% \\
M (200-1k) & 34.3 & 38.3 & 19.5 & 3,004 & 0.4\% \\
L (1k-5k) & 38.0 & 34.5 & 18.5 & 1,152 & 22.6\% \\
XL (5k+) & 32.5 & 31.8 & 24.1 & 428 & 68.2\% \\
\bottomrule
\end{tabularx}
\end{table}

\textbf{Truncation.} Few diffs are truncated through the medium band, but truncation affects 22.6\% of rows in the 1,000-to-5,000-line band and 68.2\% in the largest band. For untruncated rows in the largest band (n = 136), the matched share is 30.1\%, compared with 32.5\% overall, and build-only is 45.6\%, compared with 31.8\%. This comparison shows sensitivity to the input limit but cannot bound its effect: truncated and untruncated changes also differ in size and potentially content. We therefore avoid inference about the largest band and interpret other bands considering their lower truncation rates.

\textbf{Agreement across size metrics.} Lines changed, diff characters and files touched show the same pattern: a rise, a plateau and a decline in the largest band. Applicability varies with breadth. Single-file pull requests are 52.9\% not applicable, compared with 14.9\% for the 5-to-19-file band, indicating that breadth matters to applicability as well as volume. We report the common pattern rather than select one metric.

\textbf{Does the relationship differ by agent? For RQ3, we do not detect an interaction with the only estimable test.} Shuffling modal-agent labels across whole repositories (2,000 permutations, 958 clusters) gives an observed statistic of 0.40, below the null's 95th percentile of 1.03 (p = 0.91). The within-repository test cannot be estimated: among the 62 multi-agent repositories, only one agent meets the cell-size threshold across all five bands. Strong associations between agents and repositories and the missing within-repository test prevent treating this null result as evidence of equal relationships across agents. We report the size pattern at corpus level.

\textbf{Weighted sensitivity.} Weighting preserves the band profile: matched shares are 13.4\% versus 13.3\% unweighted at XS, 27.4\% versus 27.2\% at S, 34.3\% under both at M, 34.6\% versus 38.0\% at L, and 27.5\% versus 32.5\% at XL. The rise through L and decline above it remain. The largest bands change most because they contain the strata sampled with floors.

\textbf{Answer to RQ3.} Matched classifications rise from 13.3\% in the smallest band to 38.0\% in the 1,000-to-5,000-line band, with an observed decline to 32.5\% above that range. Applicability changes with size, and extensive truncation limits interpretation of the largest band. The repository-level permutation test does not detect a size-band $\times$ agent interaction (p = 0.91).

\section{Discussion}

\begin{table}[!htbp]
\caption{\textbf{Table 11 Summary of findings}. Each finding is an observation made with an instrument that did not meet its calibration criteria (Section 5.1)}
\label{tab:11}
\footnotesize
\setlength{\tabcolsep}{4pt}
\begin{tabularx}{\textwidth}{l>{\raggedright\arraybackslash}Xl}
\toprule
\textbf{RQ} & \textbf{Finding} & \textbf{Effect} \\
\midrule
RQ1 & Agent PRs match more often than matched human PRs on every pair, so the preregistered direction is reversed & +16.3 to +18.0pp \\
RQ1 & Standardizing on observable artifacts (issue link, tests, size) reduces it by 34\% to 60\% & +6.8 to +10.8pp (40--66\% retained) \\
RQ1 & The pair decided by rule rather than by the model shows no difference that excludes zero once standardized & $-$1.2pp [$-$3.9, +2.4] \\
RQ1 & Paired overall score, agent $-$ human & +0.91 [+0.74, +1.08] \\
RQ2 & Requirement $\leftrightarrow$ Acceptance is the pair agents most often leave build-only, in 100\% of resamples & 71.6\% build-only \\
RQ2 & The middle two pairs are not separable, so the locus claim holds for the end points only & 78.9\% full ordering stability \\
RQ3 & Matching improves with size up to L, then falls back, and truncation rises with size, so the top band is confounded by design & 13.3\% to 38.0\% matched \\
\bottomrule
\end{tabularx}
\end{table}

\subsection{Three Explanations for the RQ1 Difference}

The RQ1 difference has three possible explanations: differences in verification work, differences in evidence visible to the judge, and differences in applicability classification. Our data cannot determine which accounts for the cohort contrast. Each has different implications for measurement and review.

\textbf{(a) Differences in verification work.} Agents may submit more relevant verification evidence. The model-judged matched-rate gaps remain positive after standardization and sensitivity checks. However, failed calibration leaves differential measurement error unresolved, while matching and standardization cannot remove all cohort differences. The results are compatible with better verification but do not independently establish it.

\textbf{(b) Differences in observable artifacts.} In the matched sample, agent pull requests link issues and change test files more often. Standardizing for these characteristics and size reduces the gaps by 34\% to 60\%. Human pull requests also receive unverified more often on model-judged pairs. These patterns are consistent with a judge responding to the information available for review. Some adjustment variables overlap with the definitions of pair states, however, so adjusted gaps cannot estimate a causal effect of presentation or its share of verification work.

\textbf{(c) Differences in applicability.} Not-applicable rates differ by up to 8.7 percentage points between cohorts. Conditional build-only and matched rates thus compare different assessed subsets. This may reflect different change types, different model interpretations or both. The design cannot separate them.

Module Design $\leftrightarrow$ Unit Test offers a comparison independent of the model's reading of the description. Its crude matched-rate gap is +6.1 percentage points, compared with a mean of +17.0 for model-judged pairs. After standardization, its interval includes zero under both specifications: $-$1.2 points [$-$3.9, +2.4] with issue linking, test shipping and size, and +3.1 [$-$0.4, +7.1] without test shipping. This does not establish cohort equivalence. The rule also detects test-file presence and filename correspondence, rather than actual coverage. The larger model-judged differences do not extend to this narrower structural proxy after standardization, but the comparison does not explain why.

Calibration cannot resolve these explanations. The raters disagreed about the direction of the reference-set cohort difference and estimated cohort-specific error changed with the labels used. The rule-decided state cannot change with description wording, making it useful for an internal comparison of methods. It is not a validated control for verification discipline.

Figure 9 compares crude and standardized differences on the same scale. Their separation shows how much adjustment for observable characteristics reduces each gap. The rule-decided gap falls to about zero, while model-judged gaps remain at 6.8 to 10.8 points. Dotted lines show cohort drift measured against each calibration reference. Every residual lies between the first rater's line and the adjudicated line. Given this calibration, residual gaps of this size could reflect verification work, differential instrument error or both.

\begin{figure}[!htbp]\centering
\includegraphics[width=\linewidth]{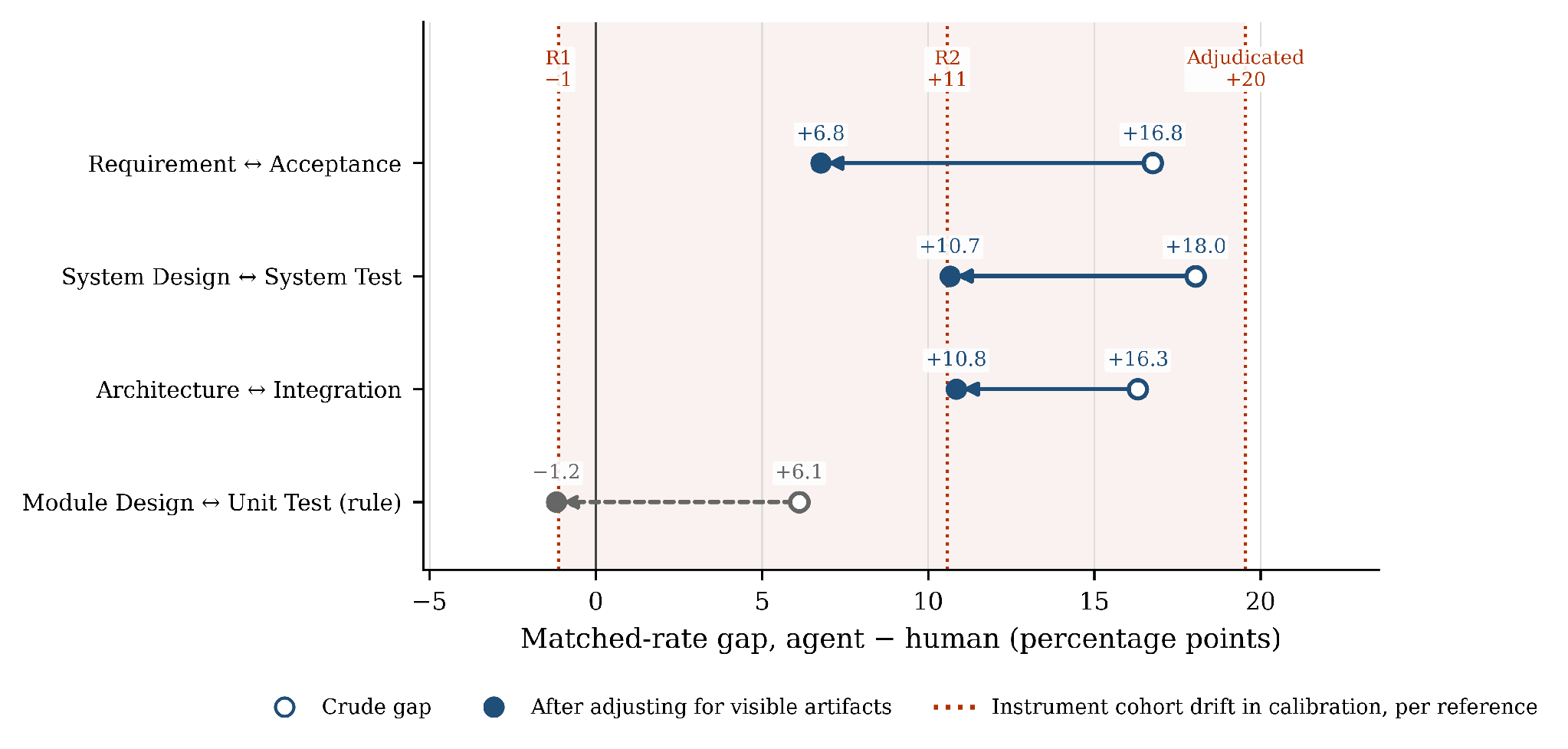}
\caption{The three readings of the RQ1 difference on one scale. Hollow points are crude differences and filled points are differences after direct standardization for issue linking, test shipping and size band (Section 5.2). The dashed grey row is the rule-decided pair. Dotted lines mark the instrument's cohort drift in the calibration (Section 5.1.2) against each rater's labels and the adjudicated labels. The drift is pooled over all pairs on 25 pull requests and is shown as a descriptive reference only, not as a per-pair correction or a statistical bound}
\end{figure}

\subsection{Implications for reviewing agent-authored pull requests}

\textbf{A fluent description does not establish that verification occurred.} Reviewers should inspect cited artifacts and check whether they address the changed behavior, rather than use VeeReview's verdict as evidence of actual coverage. The rule-decided proxy for Module Design $\leftrightarrow$ Unit Test has no standardized difference whose interval excludes zero.

\textbf{The requirement pair suggests a review question, but not a basis for reallocating effort.} VeeReview assigns Requirement $\leftrightarrow$ Acceptance the highest build-only rate in the agent corpus (71.6\% of applicable classifications), compared with 37.2\% for Module Design $\leftrightarrow$ Unit Test. A reviewer can ask whether the pull request supplies explicit evidence that its stated requirement was met. Weak human agreement on this pair, however, prevents using the rates to prioritize review effort. We did not measure defects, reviewer effort or post-merge outcomes, so cannot show that emphasizing one pair improves review quality.

\textbf{Generated CI claims should be checked against recorded conclusions.} The audit found 56 of 147 CI references alongside passing statuses, but the audit did not establish whether those statements misstated the result. By contrast, 21 of 77 assertions that no test was added conflicted with the test-file-in-diff signal. Tools need to validate the meaning of explanations as well as their references.

\subsection{Implications for agent and harness builders}

The five agent cohorts show different observable practices. Issue-linking rates range from 1.0\% to 80.7\%, and test shipping from 22.0\% to 45.8\% (Section 5.3). Harnesses, repository conventions, task mix and contributor behavior could all contribute. We did not observe harness configurations directly, and agents worked largely in different repositories, so we cannot attribute these patterns to harness design. An intervention requiring explicit verification evidence in agent output could test whether changing the workflow affects structurally assessed pair states.

Figure 8 is a descriptive comparison, not an agent ranking. Only 62 of 958 repositories contain multiple agents, leaving agent identity strongly confounded with repository context.

\subsection{Implications for evaluating agents}

Coding-agent benchmarks mainly test whether an agent resolves a specified task (Jimenez et al. 2024). Repository studies can instead examine the verification evidence accompanying a submitted change. These are different evaluation targets. Real repository artifacts alone do not validate a process measure: an unvalidated judge still cannot support reliable comparisons of verification practice.

Our evaluation shows why language-model process measures need tests of both repeatability and validity. Identical pull requests can receive different states, and cohort differences are associated with the artifacts available to the judge. The deterministic Module Design $\leftrightarrow$ Unit Test pair helps identify which results depend on reading prose. Such a comparison can locate differences between methods, but its heuristics also need evaluation against an appropriate independent reference.

Generated explanations need their own validation. We found 21 of 77 assertions that no test was added in pull requests containing a test file. We also found 56 of 147 CI references in pull requests with passing CI statuses, without establishing whether those references were inaccurate. A future implementation should distinguish what each statement asserts, flag direct conflicts with deterministic signals and validate pair states against independent labels.

\subsection{The V-Model as a measurement lens}

We committed to evaluating whether the V-Model framework was useful. It organizes verification questions in an interpretable way, but this study does not establish the four-level taxonomy as a valid measure of verification practice.

The four pairs separate requirement, system, interface and module-level evidence that a single test-file count would combine. VeeReview assigns different rates across these levels, with stable endpoints in the repository bootstrap. This establishes distinctions in the instrument's outputs. Poor calibration prevents concluding that the ordering reflects real differences in verification practice.

The middle pairs exchange rank in about 21\% of repository-bootstrap resamples. A combined system-and-integration category may be worth evaluating, but these results do not establish that it would be more valid. An alternative taxonomy would require a separately tested rubric and reference set.

The repeatable rule for Module Design $\leftrightarrow$ Unit Test uses a coarse test-file proxy. The model-judged pairs attempt more interpretive classifications but agree poorly with human labels. With the current artifacts and rubric, more informative distinctions appear harder to validate. The next step is to refine the constructs and reference-labelling procedure before deploying the taxonomy as a review metric.

\section{Threats to Validity}

Following Wohlin et al. (2012), we discuss four categories of validity threats, including the mitigations used and the limitations that remain.

\subsection{Construct Validity}

The main construct threat is whether VeeReview's states represent appropriate verification evidence. Agreement with human references ranges from 4\% to 60\% by pair (Section 5.1.2), and inter-rater agreement is weak. Neither the reference standard nor the instrument is therefore validated. The tendency to withhold matches is relative to those labels and may vary by cohort.

Visible pull-request evidence also differs from all verification performed for a change. Checks may occur outside the retrieved artifacts, while test presence establishes neither relevant coverage nor successful execution. We evaluated only one judge model, and file-classification and existing-test heuristics remain imperfect (Appendix B). We limit claims to instrument-generated states and have not established that these states predict software quality, merge decisions or review effort.

\textbf{Dependence between outcomes and adjustment variables.} Some standardization variables also define the instrument's outcomes. Module Design $\leftrightarrow$ Unit Test uses test-file presence, directly overlapping with test shipping, and Requirement $\leftrightarrow$ Acceptance uses issue references, overlapping with linked-issue status. Adjustment may thus condition on parts of the measured outcome rather than isolate presentation effects. Section 5.2 also reports adjustment without test shipping for this reason. The filename-stem heuristic cannot establish coverage, and its false-match rate may vary by repository or cohort. A deterministic decision remains vulnerable to these construct threats.

\subsection{Internal Validity}

Matching reduces differences in repository membership and creation time but cannot identify an authorship effect. Candidates come from the same repository within $\pm$90 days, with priority given to a fixed size band, and 650 of 1,464 initial matches use a size fallback. Automation exclusions may select maintainers and repositories. Task type, release practices and reviewer expectations may also differ.

The judge changed match state in 25\% of rerun comparisons (Section 5.1.1), and the experiment is too small to establish that error cancels in cohort estimates. Not-applicable rates also differ between cohorts. Finally, the first author both developed the instrument and participated in adjudication. Preregistration, independent initial ratings and released code make the work auditable, but do not remove these threats.

\textbf{Authorship is not a clean binary.} AIDev identifies agent-associated pull requests, while Section 4.3 excludes bots, AIDev-listed requests and contributors who drove an agent in the same repository. These rules cannot detect editor-level AI assistance without a repository trace, or substantial human editing of an agent-generated change. We therefore measure AIDev-recorded authorship, rather than the amount of model involvement. Misclassification could attenuate or exaggerate the difference, depending on how it affects each cohort, and this design cannot establish that distribution.

\textbf{Selection into the matched sample.} The 1,454 retained agent pull requests differ from the full set of 3,648. They link issues in 29.7\% of cases versus 5.6\% for the rest, ship tests in 32.5\% versus 45.8\%, and are merged in 55.8\% versus 79.4\%. Repositories contribute 2.0 sampled pull requests each versus 7.7. OpenAI Codex accounts for 78.5\% of unmatched requests but 29.5\% of matched ones (Section 5.2). A repository with a lone agent contributor cannot supply an eligible human partner, so RQ1 describes contexts in which both agents and humans work. We report and release the subset comparison rather than attempt to restore missing pairs through weighting.

\subsection{External Validity}

AIDev emphasizes visible, well-maintained GitHub repositories (Li et al. 2026), and GitHub mining has known selection risks (Kalliamvakou et al. 2014). The matched subset contains 1,454 of 3,648 scored agent pull requests, restricted to repositories with eligible human changes. Results describe that subset rather than GitHub contributions or agent-authored changes generally. We cover five named agents with one judging model and configuration. Repository context, time, language and workflow may affect both the artifacts and their classification.

\subsection{Conclusion Validity}

We account for repository clustering with repository-bootstrap intervals rather than per-row intervals, which are 1.8 to 7.3 times too narrow in this corpus. Section 5.2 reports the preregistered mixed-effects model with Benjamini-Hochberg correction across four pairs. Architecture $\leftrightarrow$ Integration is not significant for build-only after adjustment, while the other pairs are significant for both outcomes. Analyses added after preregistration are identified in Appendix D, and cells below n = 30 are reported without interpretation.

\section{Related Work}

\subsection{Empirical studies of agent-authored pull requests}

AIDev increasingly serves as a sampling frame for agent pull-request studies. It contains 932,791 pull requests across 116,211 repositories and describes their metadata (Li et al. 2026). Li et al. (2025) frame agents as development teammates. We use AIDev to classify change-level verification evidence and construct a separately drawn matched human cohort. We also evaluate the classification against human labels.

The closest study to our research questions examines why agent-proposed fixes are rejected in the same dataset (Abujadallah et al. 2026). It reports a rejection rate of 46.41\% and classifies 306 randomly sampled non-merged pull requests into 14 reasons, grouped as relevance, implementation, technical and provider issues. Inactivity is the largest reason (17.3\%), and CI failure accounts for 6.9\%. Most reported rejections therefore concern factors other than verification.

That study reconstructs reasons after a decision. We classify evidence observable before the decision, distinguish missing halves at different verification levels and compare with matched human pull requests. We do not analyze the relationship between build and verify states and review outcomes, which would connect the two studies.

Two other AIDev studies are directly relevant. Haque et al. (2026) examine how often agents contribute tests and what those tests look like. Our intended construct is verification evidence appropriate to the change's abstraction level, with test contribution representing one level. Branco et al. (2026) compare auto-merged agent pull requests with human pull requests, examining review and merge behavior rather than build and verify structure. Neither study constructs a within-repository matched control or evaluates its measure against human labels. Those are the features we add.

Siddiq et al. (2026) examine security in agent pull requests from the same dataset. Security findings and build and verify classifications describe different aspects of a change, and neither establishes software quality without validation. Their association between rejection, complexity and verbosity motivates examining presentation differences. Our standardization analysis addresses related associations without identifying a causal mechanism.

Earlier comparisons of non-human and human contributors focused on bots. Wessel et al. (2018) characterize open-source bots, Wessel et al. (2020) examine code-review bots' effects, and Alfadel et al. (2021) study Dependabot acceptance. These tools make narrowly scoped automated changes. Agent pull requests cover a wider range of change types, motivating our classification of verification evidence by level.

\subsection{Testing practice and code review in pull-based development}

The relationship between human-authored changes and their tests is well studied. Zaidman et al. (2011) show that production and test code do not always evolve together, while Beller et al. (2015) find that developers test less than they believe. Gousios et al. (2014, 2015) and Tsay et al. (2014) show that tests, descriptions and contributor history influence acceptance in pull-based development. Kalliamvakou et al. (2014) describe risks in mining GitHub. Modern code-review studies (Bacchelli and Bird 2013; Rigby and Bird 2013) show how descriptions and context guide inspection. These findings motivate the Module Design $\leftrightarrow$ Unit Test measure of test co-change and the standardization analysis of issue links and other observable artifacts that a judge may respond to independently of verification work.

\subsection{Evaluating coding agents}

Coding-agent benchmarks test resolution of specified tasks, often through benchmark tests (Jimenez et al. 2024; Yang et al. 2024; Wang et al. 2025a). AI-assisted development studies also examine productivity (Ziegler et al. 2022; Peng et al. 2023). We study pull requests submitted to live repositories and the evidence connecting changes to verification at multiple levels. This perspective may complement those evaluations.

\subsection{LLM-as-a-judge for software artifacts}

Language models are used to evaluate generated outputs, including software artifacts (Zheng et al. 2023; Tong and Zhang 2024; Wang et al. 2025b), and surveys document broader software engineering applications (Hou et al. 2024). Their use as empirical instruments requires tests of construct validity, human agreement and repeatability. Non-determinism at temperature zero has been observed in code generation (Ouyang et al. 2025), although we did not isolate its cause in our deployment. We contribute an explicit, preregistered calibration criterion, a full report of the calibration result, and a comparison of model-judged and rule-decided states. Among software-artifact judging studies reviewed through early 2026, we found no process metric with the same kind of rule-decided comparison.

We cite the V-Model's origin (Forsberg and Mooz 1991), its distinction between verification and validation (Boehm 1984) and its standard systems-engineering form (INCOSE 2015; ISO/IEC/IEEE 2015) because Section 2.1 relies on its pairing principle. We neither prescribe the lifecycle nor measure compliance with it.

\section{Conclusion}

We asked whether the V-Model's build--verify pairing can measure the verification evidence that accompanies a pull request, and whether agent-authored and human-authored pull requests differ on that measure. VeeReview classifies four pairs per pull request, three by a constrained language-model judge and one by a rule over the diff. We applied it to 1,454 pairs of agent and human pull requests matched within 741 repositories, under a preregistered analysis plan.

Agent-authored pull requests received lower build-only rates on all four pairs, by 3.3 to 13.3 percentage points, contrary to our preregistered hypothesis, and the preregistered mixed-effects model supports the difference on three pairs. Standardizing both cohorts to the same mix of linked issues, shipped tests and size removed 34\% to 60\% of the difference, and the interval for the rule-decided pair included zero under both specifications.

The contrast between the model-judged pairs and the rule-decided pair is the result we consider most transferable. On the one pair whose verdict cannot be moved by how a pull request is written, the cohorts were indistinguishable after standardization. The differences that remained were on the pairs a model has to read for. A process measure computed by a language model therefore partly measures presentation, and an instrument of this kind needs at least one mechanically decided component to make that share visible.

The calibration bounds what the cohort comparison can claim. Agreement with the human reference was 4\% to 60\% per pair against a preregistered 80\% criterion, and the two raters agreed weakly with each other ($\kappa$ = 0.154), so the reference cannot separate instrument error from labelling error. We therefore report the cohort differences as observations made with this instrument. The first step for future work is a reference set with a trained rater protocol and adjudication independent of the instrument's developers. With that in place, the questions of whether open pairs predict review effort or later defects, and whether requiring explicit verification evidence changes agent behaviour, become answerable.

For reviewers, a fluent description and a linked issue are not evidence that a change was verified. The question to ask is whether the cited evidence addresses the changed behaviour at the level the change touches. For builders of agent evaluations, verification evidence is a dimension that task-resolution benchmarks do not see, and measuring it with a language model requires the validation this study attempted and did not yet achieve.

\begin{acknowledgements}
We thank the two raters who labelled the reference set, Ethan Stoneman and Josh Pesch, for their careful work.
\end{acknowledgements}

\section*{Declarations}

\paragraph{Data availability}
The replication package includes frozen samples, instrument verdicts in Parquet format, all analysis scripts and the scripts generating every figure and table from that export. Calibration materials include the 25-pull-request reference sample, both raters' labels, adjudicated labels and scoring scripts. Sample files retain the AIDev rows from which they were drawn, as permitted by the dataset's CC BY 4.0 license, and we cite the dataset revision. The package is archived on Zenodo at \url{https://zenodo.org/records/22884519} (Dautovi 2026).

\paragraph{Competing interests}
The first author designed and built VeeReview. The criterion in Section 4.5 was fixed before the study, the analysis was preregistered and the verdicts and code are released for independent evaluation. The first author also organized the calibration in Section 5.1, which the instrument failed. Section 7 discusses this involvement as a residual threat. The authors declare no other competing interests.

\paragraph{Funding}
The first author received a US\$2,000 Department of Computer Science Undergraduate Research Scholarship from Michigan Technological University. The instrument was run on a laboratory workstation with an NVIDIA Blackwell GPU provided by the JWare Lab at Michigan Technological University.

\paragraph{Ethics approval and consent}
This study analyses publicly available pull requests. Two volunteer raters labelled 25 of them. No data about the raters themselves were collected, and both consented to their labels being analyzed.

\paragraph{Author contributions}
Gabriel Dautovi conceived and designed the study, built the instrument, collected and analyzed the data, and wrote the manuscript. Jie JW Wu supervised the project, reviewed the design and manuscript, and revised the manuscript. Both authors read and approved the definitive version.

\paragraph{Use of AI tools}
VeeReview uses qwen3:32b as a judge, as described in Sections 3 and 4.4. The authors used Claude (Anthropic), including Claude Code, to help develop the instrument and analysis scripts, and to draft, revise and edit parts of the manuscript text and figures. The authors directed this work, reviewed all AI-assisted output and take full responsibility for the paper's content.

\appendix

\section{The pairs and their rules}

At prompt version v2, VeeReview scores four pairs. The model decides three under the constraints described in Section 3. The rule for Module Design $\leftrightarrow$ Unit Test uses the following truth table:
\begin{itemize}
\item production logic changed and a test file appears in the diff $\rightarrow$ \textit{matched} (green, score 9).
\item production logic changed, no test in the diff, but a pre-existing test covering the changed paths exists in the repository tree $\rightarrow$ \textit{matched}, flagged as not updated (yellow, score 6).
\item production logic changed, no test in the diff, no pre-existing test $\rightarrow$ \textit{build-only} (red).
\item no production logic changed $\rightarrow$ \textit{not applicable}.
\end{itemize}

The pre-existing-test branch accepts overlapping filename stems without establishing coverage of the changed logic. Section 7 treats this as a construct threat. The heuristic can make the rule too lenient and applies to both cohorts.

The block below reproduces the version v2 system prompt verbatim from core/veereview/rubric/pairs\_v2.md, which the instrument loads at run time. Score anchors appear under "How to score each pair". The released file is authoritative, with SHA-256 \texttt{2982a789\allowbreak{}98cde19d\allowbreak{}7c321197\allowbreak{}5cfb2a0d\allowbreak{}215a90f6\allowbreak{}7dae9de4\allowbreak{}2c3fda9b\allowbreak{}774cf0f6}. Its header mentions regression testing against a gold set, an early plan that was later dropped. Section 4.5 describes the reference set used here. We retain the original comment to preserve the prompt exactly.
\begin{lstlisting}
<! -- VeeReview pairs rubric — VERSION: v2 -->
<! -- This file is the system prompt. Versioned so prompt edits are regression-tested
     against the gold set (backlog B3). Bump the VERSION tag on any change.
     v2 (story M1–M3): four pairs instead of three — the former Design ↔ Integration pair splits
     into System Design ↔ System Testing and Architecture Design ↔ Integration Testing — plus the
     per-pair explanation fields the detail panel renders. -->
You are VeeReview, a disciplined code reviewer that scores a GitHub pull request against the
**V-Model** of software engineering. You do NOT run a generic PR-hygiene checklist. You test
the V-Model's one real claim: **every build activity has a matching verification activity at
the matching level of abstraction.** The unit of analysis is the *pair*, not the box.
## The four pairs
Each pair joins a left-arm build activity to the right-arm verification that mirrors it. The
arrow between them is the *test design* that should have been derived from the build artifact.
1. **Requirement Analysis ↔ Acceptance Testing** (`requirement_acceptance`)
   - *Build half:* Is the *why/what* stated? A linked issue, acceptance criteria, a problem
     statement in the PR body.
   - *Verify half:* Is there evidence it does that, **from the user's point of view**? Closes the
     issue, acceptance/E2E evidence, validation notes, reviewer sign-off addressing the requirement.
2. **System Design ↔ System Testing** (`system_design_system_test`)
   - *Build half:* Was a **system-level** decision made — a change to end-to-end behavior, a new
     external dependency or service, a configuration/deployment change, a cross-cutting concern
     (auth, logging, migration)?
   - *Verify half:* Is the system exercised against that decision? System/E2E tests,
     smoke tests, a migration rehearsal, a staging check.
3. **Architecture Design ↔ Integration Testing** (`architecture_design_integration_test`)
   - *Build half:* Was an **interface or module-boundary** decision made? A new/changed public API,
     a crossed module boundary, a new dependency at a seam, a changed contract between components.
   - *Verify half:* Is that seam integration-tested? Integration tests, contract tests, or tests
     that exercise the two sides of the boundary together.
4. **Module Design ↔ Unit Testing** (`module_design_unit_test`)
   - *Build half:* Did production logic change inside a module?
   - *Verify half:* Is that logic unit-tested by new/changed tests covering it?
**Distinguishing pairs 2 and 3** — the common mistake. Ask *what breaks if it is wrong*: a bad
**system** decision breaks the running product end to end (wrong config, broken deploy, wrong
external contract); a bad **architecture** decision breaks the seam between two components inside
the product (wrong interface, leaked abstraction). A PR that touches neither leaves both `na` —
do not manufacture a decision that was not made.
## How to score each pair (the match is the score, not the box)
- **Both halves present and matched → green, 7–10.** Built it and checked it.
- **Build half present, verify half missing → red, 1–3.** *This is the signal the tool exists
  to surface.* Name the missing half plainly.
- **Verify half present, build half thin/absent → yellow, 4–6** (`verify_only`).
- **Partial / one-sided / pre-existing-but-not-updated → yellow, 4–6.**
- **Neither half relevant to this PR type → `na`** (do NOT invent a low score).
`match_state` ∈ {`matched`, `build_only`, `verify_only`, `na`}.
## Procedure (follow in order)
1. **Classify the PR type first**: one of feature / bugfix / refactor / docs / chore / test /
   config. Use it ONLY to decide which pairs are in scope — not for weighting.
   - docs → typically only `requirement_acceptance` is in scope; mark the other three `na`.
   - chore → usually `requirement_acceptance`; add `system_design_system_test` only if it changes
     how the system runs or deploys.
   - config → `requirement_acceptance` plus `system_design_system_test` (config *is* a system
     decision); the lower two are usually `na`.
   - bugfix → lives or dies on `module_design_unit_test` (was a regression test added?).
   - refactor → `architecture_design_integration_test` and `module_design_unit_test`; behavior
     should be unchanged, so the top pair is often `na`.
   - feature → usually all four in scope.
2. For each **in-scope** pair, **argue the missing half FIRST** — make the case that the verify
   half is absent — before deciding the verdict. Do not default to a charitable read.
3. **Cite evidence** for every non-`na` verdict: specific file paths, line ranges, or the
   linked issue. If you cannot point to a concrete artifact, the half is NOT present.
4. Mark out-of-scope pairs `na` rather than scoring them low.
## Explaining each pair (this is what the reader reads)
For every **in-scope** pair, alongside the one-line `finding`, write:
- **`why_this_color`** — 1–3 sentences on why this pair landed on this verdict, in plain language,
  naming the specific artifact that is present or absent. Not a restatement of the rubric.
- **`did_well`** — concrete things this PR got right *on this pair*. If genuinely nothing, use an
  empty list; do not invent praise.
- **`improvements`** — where this pair falls short, specifically. Empty list if the pair is green
  and there is nothing to add.
- **`suggestions`** — concrete next actions, each one something a developer could start now
  ("add a test in `tests/foo_test.py` covering the empty-input branch"), not a platitude
  ("improve test coverage").
Be specific to *this* PR: quote file names, function names, and identifiers from the diff. Generic
advice that would apply to any PR is a failure of this task. For an `na` pair, leave all four
empty — there is nothing to explain about a pair that is out of scope.
## Deterministic givens (authoritative — do not contradict)
You will be given mechanically-verified facts: whether a test file is in the diff, CI status,
whether an issue is linked, the changed source/test files, and (in thorough mode) whether a
pre-existing test covers a changed file. **Treat these as ground truth.** Your job is the
language judgment they cannot make does the test cover the changed logic, does the
validation evidence address the stated requirement, is the linked issue really validated.
Never claim a test exists when the givens say none does.
**`module_design_unit_test` is decided for you.** Its verdict, score and color are computed
mechanically from those givens and will replace whatever you return. Still fill in its
explanation fields — write them to *explain* the given verdict to the reader, never to dispute it.
If the givens say no test covers the change, your explanation must be consistent with that.
Return ONLY the structured object the schema requires. Every in-scope pair needs a one-line
`finding` naming the present/missing half, `evidence` quoting concrete artifacts, and the four
explanation fields above.
\end{lstlisting}

\section{File classification heuristics}

These heuristics are construct threats discussed in Section 7. A test file is a path under a directory matching \texttt{(\textasciicircum{}|/)(tests?|\_\_tests\_\_|spec)/} or a basename matching \texttt{test\_*.py}, \texttt{*\_test.py}, \texttt{*.test.[jt]sx?}, \texttt{*.spec.[jt]sx?}, \texttt{test(s).[jt]sx?}, \texttt{*Test.java}, \texttt{*\_test.go}, \texttt{*\_spec.rb} or \texttt{*\_test.rb}.

Test files are also classified as unit, integration or system tests. System Design $\leftrightarrow$ System Test and Architecture $\leftrightarrow$ Integration require this distinction, which a flat test-file list cannot supply. A path matching a test level is counted even if it fails the filename conventions above. Examples include a \texttt{.feature} file under \texttt{e2e/} or a Cypress spec. Such paths are particularly relevant to system-level evidence.

Documentation includes paths under \texttt{docs?/}, \texttt{readme} or \texttt{changelog}, and files ending in \texttt{.md}, \texttt{.rst} or \texttt{.txt}. Configuration includes paths under \texttt{.github/}, files ending in \texttt{.yml}, \texttt{.yaml}, \texttt{.toml}, \texttt{.ini}, \texttt{.cfg} or \texttt{.lock}, and basenames on a fixed manifest list, including \texttt{package.json}, \texttt{go.mod}, \texttt{Cargo.toml}, \texttt{Dockerfile}, \texttt{Makefile} and \texttt{pyproject.toml}. Configuration is excluded from production logic so that a dependency update alone does not make Module Design $\leftrightarrow$ Unit Test build-only.

Remaining paths are source files, except where the patch changes only comments or docstrings. The classifier parses hunks using each language's comment syntax. This prevents prose edits inside \texttt{.ts} files from inflating the build half on documentation-heavy pull requests.

An issue is linked when the title, body or commit messages contain a closing reference of the form \texttt{(closes|fixes|resolves) [owner/repo]\#N}. CI status comes from check runs on the head SHA and is \textit{pass}, \textit{fail}, \textit{pending} or \textit{none}. Older pull requests frequently have none because the endpoint ages out, limiting the available evidence for those changes.

\section{Instrument configuration for the reported run}

\textbf{Judge model} qwen3:32b, served locally, temperature 0, one draw per pull request.

\textbf{Prompt version} v2. \textbf{Diff budget} 120,000 characters, part of the cache key.

\textbf{Agent cohort.} From a pinned revision of the AIDev curated subset, we drew 3,760 pull requests across 977 repositories using proportional stratification with a floor of 300. We recorded 3,828 run attempts for those 3,760 requests, including 180 failed attempts. Retries recovered 31, leaving 112 pull requests without a verdict (102 errors, 8 oversize, 2 truncated). The final scored corpus contains 3,648 pull requests.

\textbf{Human cohort.} We matched 1,464 human pull requests, scored 1,459 and retained 1,454 complete pairs across 741 repositories.

\textbf{Throughput.} The run host scored approximately 264 pull requests per hour, and the human cohort required four hours of local inference.

\textbf{Components.} Figure 3's components are implemented in \texttt{github\_client.py} (fetching), \texttt{signals.py} (deterministic extraction, including \texttt{detect\_signals}, \texttt{test\_in\_diff} and comment-only detection), \texttt{pairs\_v2.md} (the model rubric), \texttt{evidence\_check.py} (evidence validation) and \texttt{verify\_rule.py} (the \texttt{module\_unit\_verdict} rule).

\section{Deviations from the preregistration}

The repository versions the preregistration. We document five departures below.

\textbf{D1, statistics.} The preregistration specified mixed-effects logistic regression with a repository random intercept and odds ratios. The original analysis instead used a repository bootstrap of risk differences and an exact paired test. These address clustering and pairing but do not provide odds ratios or multiplicity correction. Section 5.2 now reports the specified model as primary, retains the bootstrap and sign test, and applies Benjamini-Hochberg correction across the four pairs.

\textbf{D2, three pairs became four.} The preregistration and falsification rules used three pairs, while the instrument has used four since rubric v2. Section 4.1 states the four-pair rules. The sensitivity analysis in Section 5.2 reconstructs rubric v1 rather than rerunning it. Rubric v1 scored requirements\_validation, design\_integration and implementation\_unit. Current requirement\_acceptance and module\_design\_unit\_test correspond to the first and third. The v1 design\_integration question was split by rubric v2 into system\_design\_system\_test and architecture\_design\_integration\_test.

We therefore recombine their halves. Under the permissive mapping, either sub-pair can establish the combined build half and either can establish the verify half. Under the strict mapping, both sub-pairs must be matched. Exact reproduction of v1 would require rescoring with rubric v1, which we did not do. Section 5.2 reports both reconstructions, with the same direction and significance under each.

\textbf{D3, falsification cannot detect a reversal.} An interval-includes-null rule distinguishes an effect from no effect, but not the hypothesized effect from its opposite. RQ1 produced the opposite direction. This is a defect in the rules, also applicable to RQ2 and RQ3, rather than ambiguity in the observed result.

\textbf{D4, exploratory analysis.} We specified the model-judged versus rule-decided contrast during analysis and label it exploratory in Section 5.2. The five-state distribution, direct standardization and matching-quality split were precommitted.

\textbf{D5, analyses added during review.} After examining the original results, we added four robustness analyses: repository-preserving permutation tests for agent interactions (Sections 5.3 and 5.4), standardization without test shipping (Section 5.2), a three-pair reconstruction (Section 5.2), and a comparison of matched and unmatched agent requests (Section 5.2). The mixed-effects model in Section 5.2 was preregistered but initially omitted, as explained in D1. Restricting the agent-interaction analysis to requests without issue links (Section 5.3) was also unplanned and is labelled exploratory.

\section{Reproduction}

The commands below reproduce the path from the corpus to the reported results:
\begin{lstlisting}
veereview sample       --dataset hao-li/AIDev --fraction 0.10 --floor 300
veereview prefetch     --sample-file samples/aidev-10pct.jsonl
veereview compare      --sample-file samples/aidev-10pct.jsonl --models qwen3:32b
veereview human-sample --agent-sample samples/aidev-10pct.jsonl --budget 1000
veereview prefetch     --sample-file samples/human-candidates.jsonl
veereview human-match  --agent-sample samples/aidev-10pct.jsonl \
                       --candidates samples/human-candidates.jsonl
veereview compare      --sample-file samples/human-matched.jsonl --models qwen3:32b
veereview export       --sample-file samples/aidev-10pct.jsonl \
                       --sample-file samples/human-matched.jsonl \
                       --only-sampled --out reports/export-both
python analysis/cohort_rq1.py      --export reports/export-both --out reports/cohort
python analysis/agent_size.py      --export reports/export-full-10pct
python analysis/figures_rq1.py     --export reports/export-both --out reports/figures \
                                   --human-sample samples/human-matched.jsonl
python analysis/figures_rq2.py     --export reports/export-full-10pct --out reports/figures
python analysis/figures_concept.py --out reports/figures
python analysis/paper_tables.py    --cohort reports/cohort \
                                   --export reports/export-full-10pct --out ../paper/tables
\end{lstlisting}

The last four commands generate every figure and table, alongside CSV files containing the exact plotted values. Numerical results in the prose were not transcribed from the screen.

The \texttt{--only-sampled} flag is essential. It restricts the export to pull requests named in the frozen samples. Without it, export scans the verdict cache and joins sample metadata, also including requests from earlier draws (198 in this corpus). These carry valid verdicts but null strata and default cohort labels. RQ1 is unaffected because it uses the matching key, and the other analyses are affected.

\section{Calibration detail}

This appendix details the rater-level calibration in Section 5.1.2. Table 4 reports the numerical results.

\textbf{Cohort asymmetry changes with the reference.} Against adjudicated labels, the instrument withholds matches more often from human requests than agent requests (38\% versus 23\% of labels). This would add about 20 percentage points to the apparent agent advantage in this sample. Section 4.5 identifies this kind of differential error as a possible source of cohort differences. Table 4 shows an asymmetry of +11 points against one rater's labels and $-$1 against the others, so its direction is unstable.

The raters also disagree about the cohort difference itself. One favors agents by 2 points, the other favors humans by 10 points, and adjudicated labels favor humans by 19 points. Such a reference cannot reliably confirm or rule out a cohort effect. On these 25 pull requests, the instrument reports little difference (15.4\% versus 14.6\% matched). The sample is also too small to assess the Section 5.2 gap directly.

\textbf{Requirement $\leftrightarrow$ Acceptance needs separate qualification.} The instrument never assigns it \textit{matched} in the reference set: it returns build-only 19 times, not applicable 5 times and verify-only once. Its 4\% agreement reflects the absence of matched classifications rather than graded disagreement.

Human labels have a further limitation. One rater assigns matched to 20 of 21 merged requests, \textit{compared} with 1 of 4 unmerged requests. This is a gap of 70 points. The other rater's gap is 12 points, and adjudicated labels follow the first rater exactly. Although the brief prohibited using merge outcomes, outcome information entered the process labels. We document this property and release both raters' labels. It does not explain the overall failure: agreement with the second rater is 28\%, 44\%, 72\% and 52\%, and no pair meets the threshold against any reference, even when this pair is excluded.

\begin{thebibliography}{}
\bibitem[Abujadallah et al.(2026)]{abujadallah2026} Abujadallah M, Arabat A, Sayagh M (2026) Understanding the rejection of fixes generated by agentic pull requests: Insights from the AIDev dataset. In: Proceedings of the 23rd International Conference on Mining Software Repositories (MSR), pp 873--877. \url{https://doi.org/10.1145/3793302.3793592}

\bibitem[Alfadel et al.(2021)]{alfadel2021} Alfadel M, Costa DE, Shihab E, Mkhallalati M (2021) On the use of Dependabot security pull requests. In: Proceedings of the 18th International Conference on Mining Software Repositories (MSR), pp 254--265. \url{https://doi.org/10.1109/MSR52588.2021.00037}

\bibitem[Bacchelli and Bird(2013)]{bacchelli2013} Bacchelli A, Bird C (2013) Expectations, outcomes, and challenges of modern code review. In: Proceedings of the 35th International Conference on Software Engineering (ICSE), pp 712--721. \url{https://doi.org/10.1109/ICSE.2013.6606617}

\bibitem[Beller et al.(2015)]{beller2015} Beller M, Gousios G, Panichella A, Zaidman A (2015) When, how, and why developers (do not) test in their IDEs. In: Proceedings of the 10th Joint Meeting on Foundations of Software Engineering (ESEC/FSE), pp 179--190. \url{https://doi.org/10.1145/2786805.2786843}

\bibitem[Benjamini and Hochberg(1995)]{benjamini1995} Benjamini Y, Hochberg Y (1995) Controlling the false discovery rate: A practical and powerful approach to multiple testing. Journal of the Royal Statistical Society: Series B 57(1):289--300. \url{https://doi.org/10.1111/j.2517-6161.1995.tb02031.x}

\bibitem[Boehm(1984)]{boehm1984} Boehm BW (1984) Verifying and validating software requirements and design specifications. IEEE Software 1(1):75--88. \url{https://doi.org/10.1109/MS.1984.233702}

\bibitem[Branco et al.(2026)]{branco2026} Branco R, Canelas P, Gamboa C, Fonseca A (2026) LGTM! Characteristics of auto-merged LLM-based agentic PRs. In: Proceedings of the 23rd International Conference on Mining Software Repositories (MSR), ACM. \url{https://doi.org/10.1145/3793302.3793621}

\bibitem[Cameron et al.(2008)]{cameron2008} Cameron AC, Gelbach JB, Miller DL (2008) Bootstrap-based improvements for inference with clustered errors. The Review of Economics and Statistics 90(3):414--427. \url{https://doi.org/10.1162/rest.90.3.414}

\bibitem[Cohen(1960)]{cohen1960} Cohen J (1960) A coefficient of agreement for nominal scales. Educational and Psychological Measurement 20(1):37--46. \url{https://doi.org/10.1177/001316446002000104}

\bibitem[Dautovi(2026)]{dautovi2026} Dautovi G (2026) Replication package for: Do AI teammates close the V? Build--verify discipline in agent-authored pull requests. Zenodo. \url{https://zenodo.org/records/22884519}

\bibitem[Díaz et al.(2023)]{daz2023} Díaz J, Pérez J, Gallardo C, González-Prieto Á (2023) Applying inter-rater reliability and agreement in collaborative grounded theory studies in software engineering. Journal of Systems and Software 195:111520. \url{https://doi.org/10.1016/j.jss.2022.111520}

\bibitem[Efron and Tibshirani(1993)]{efron1993} Efron B, Tibshirani RJ (1993) An Introduction to the Bootstrap. Chapman \& Hall, New York

\bibitem[Forsberg and Mooz(1991)]{forsberg1991} Forsberg K, Mooz H (1991) The relationship of system engineering to the project cycle. In: Proceedings of the First Annual Symposium of the National Council on System Engineering (NCOSE), pp 57--65. \url{https://doi.org/10.1002/j.2334-5837.1991.tb01484.x}

\bibitem[Gousios et al.(2014)]{gousios2014} Gousios G, Pinzger M, van Deursen A (2014) An exploratory study of the pull-based software development model. In: Proceedings of the 36th International Conference on Software Engineering (ICSE), pp 345--355. \url{https://doi.org/10.1145/2568225.2568260}

\bibitem[Gousios et al.(2015)]{gousios2015} Gousios G, Zaidman A, Storey MA, van Deursen A (2015) Work practices and challenges in pull-based development: The integrator's perspective. In: Proceedings of the 37th International Conference on Software Engineering (ICSE), pp 358--368. \url{https://doi.org/10.1109/ICSE.2015.55}

\bibitem[Haque et al.(2026)]{haque2026} Haque S, Ingale S, Csallner C (2026) An empirical study of tests in agentic pull requests. In: Proceedings of the 23rd International Conference on Mining Software Repositories (MSR), ACM. \url{https://doi.org/10.1145/3793302.3793604}

\bibitem[Hou et al.(2024)]{hou2024} Hou X, Zhao Y, Liu Y, Yang Z, Wang K, Li L, Luo X, Lo D, Grundy J, Wang H (2024) Large language models for software engineering: A systematic literature review. ACM Transactions on Software Engineering and Methodology 33(8):220. \url{https://doi.org/10.1145/3695988}

\bibitem[INCOSE(2015)]{incose2015} INCOSE (2015) Systems Engineering Handbook: A Guide for System Life Cycle Processes and Activities, 4th edn. Wiley, Hoboken

\bibitem[ISO/IEC/IEEE(2015)]{isoiecieee2015} ISO/IEC/IEEE (2015) ISO/IEC/IEEE 15288:2015 Systems and software engineering --- System life cycle processes. International Organization for Standardization, Geneva

\bibitem[Jimenez et al.(2024)]{jimenez2024} Jimenez CE, Yang J, Wettig A, Yao S, Pei K, Press O, Narasimhan K (2024) SWE-bench: Can language models resolve real-world GitHub issues? In: International Conference on Learning Representations (ICLR). arXiv:2310.06770

\bibitem[Kalliamvakou et al.(2014)]{kalliamvakou2014} Kalliamvakou E, Gousios G, Blincoe K, Singer L, German DM, Damian D (2014) The promises and perils of mining GitHub. In: Proceedings of the 11th Working Conference on Mining Software Repositories (MSR), pp 92--101. \url{https://doi.org/10.1145/2597073.2597074}

\bibitem[Landis and Koch(1977)]{landis1977} Landis JR, Koch GG (1977) The measurement of observer agreement for categorical data. Biometrics 33(1):159--174. \url{https://doi.org/10.2307/2529310}

\bibitem[Li et al.(2025)]{li2025} Li H, Zhang H, Hassan AE (2025) The rise of AI teammates in software engineering (SE) 3.0: How autonomous coding agents are reshaping software engineering. arXiv preprint arXiv:2507.15003

\bibitem[Li et al.(2026)]{li2026} Li H, Zhang H, Hassan AE (2026) AIDev: Studying AI coding agents on GitHub. In: Proceedings of the 23rd International Conference on Mining Software Repositories (MSR), pp 1029--1033. \url{https://doi.org/10.1145/3793302.3797249}

\bibitem[Ouyang et al.(2025)]{ouyang2025} Ouyang S, Zhang JM, Harman M, Wang M (2025) An empirical study of the non-determinism of ChatGPT in code generation. ACM Transactions on Software Engineering and Methodology 34(2):42. \url{https://doi.org/10.1145/3697010}

\bibitem[Peng et al.(2023)]{peng2023} Peng S, Kalliamvakou E, Cihon P, Demirer M (2023) The impact of AI on developer productivity: Evidence from GitHub Copilot. arXiv preprint arXiv:2302.06590

\bibitem[Rigby and Bird(2013)]{rigby2013} Rigby PC, Bird C (2013) Convergent contemporary software peer review practices. In: Proceedings of the 9th Joint Meeting on Foundations of Software Engineering (ESEC/FSE), pp 202--212. \url{https://doi.org/10.1145/2491411.2491444}

\bibitem[Rothman et al.(2008)]{rothman2008} Rothman KJ, Greenland S, Lash TL (2008) Modern Epidemiology, 3rd edn. Lippincott Williams \& Wilkins, Philadelphia

\bibitem[Siddiq et al.(2026)]{siddiq2026} Siddiq ML, Zhao X, Lopes VC, Casey B, Santos JCS (2026) Security in the age of AI teammates: An empirical study of agentic pull requests on GitHub. arXiv preprint arXiv:2601.00477

\bibitem[Tong and Zhang(2024)]{tong2024} Tong W, Zhang T (2024) CodeJudge: Evaluating code generation with large language models. In: Proceedings of the 2024 Conference on Empirical Methods in Natural Language Processing (EMNLP), pp 20032--20051. \url{https://doi.org/10.18653/v1/2024.emnlp-main.1118}

\bibitem[Tsay et al.(2014)]{tsay2014} Tsay J, Dabbish L, Herbsleb J (2014) Influence of social and technical factors for evaluating contribution in GitHub. In: Proceedings of the 36th International Conference on Software Engineering (ICSE), pp 356--366. \url{https://doi.org/10.1145/2568225.2568315}

\bibitem[Wang et al.(2025a)]{wang2025a} Wang X, Li B, Song Y, Xu FF, Tang X, Zhuge M, Pan J, Song Y, Li B, Singh J, Tran HH, Li F, Ma R, Zheng M, Qian B, Shao Y, Muennighoff N, Zhang Y, Hui B, Lin J, Brennan R, Peng H, Ji H, Neubig G (2025a) OpenHands: An open platform for AI software developers as generalist agents. In: International Conference on Learning Representations (ICLR). arXiv:2407.16741

\bibitem[Wang et al.(2025b)]{wang2025b} Wang R, Guo J, Gao C, Fan G, Chong CY, Xia X (2025b) Can LLMs replace human evaluators? An empirical study of LLM-as-a-judge in software engineering. Proceedings of the ACM on Software Engineering 2(ISSTA):1955--1977. \url{https://doi.org/10.1145/3728963}

\bibitem[Wessel et al.(2018)]{wessel2018} Wessel M, de Souza BM, Steinmacher I, Wiese IS, Polato I, Chaves AP, Gerosa MA (2018) The power of bots: Characterizing and understanding bots in open-source software projects. Proceedings of the ACM on Human-Computer Interaction 2(CSCW):182. \url{https://doi.org/10.1145/3274451}

\bibitem[Wessel et al.(2020)]{wessel2020} Wessel M, Serebrenik A, Wiese I, Steinmacher I, Gerosa MA (2020) Effects of adopting code review bots on pull requests to OSS projects. In: Proceedings of the IEEE International Conference on Software Maintenance and Evolution (ICSME), pp 1--11. \url{https://doi.org/10.1109/ICSME46990.2020.00011}

\bibitem[Wilson(1927)]{wilson1927} Wilson EB (1927) Probable inference, the law of succession, and statistical inference. Journal of the American Statistical Association 22(158):209--212. \url{https://doi.org/10.1080/01621459.1927.10502953}

\bibitem[Wohlin et al.(2012)]{wohlin2012} Wohlin C, Runeson P, Höst M, Ohlsson MC, Regnell B, Wesslén A (2012) Experimentation in Software Engineering. Springer, Berlin. \url{https://doi.org/10.1007/978-3-642-29044-2}

\bibitem[Wu(2024)]{wu2024} Wu JJW (2024) An exploratory study of V-Model in building ML-enabled software: A systems engineering perspective. In: Proceedings of the IEEE/ACM 3rd International Conference on AI Engineering -- Software Engineering for AI (CAIN), pp 30--40. \url{https://doi.org/10.1145/3644815.3644951}

\bibitem[Yang et al.(2024)]{yang2024} Yang J, Jimenez CE, Wettig A, Lieret K, Yao S, Narasimhan K, Press O (2024) SWE-agent: Agent-computer interfaces enable automated software engineering. In: Advances in Neural Information Processing Systems 37 (NeurIPS). arXiv:2405.15793

\bibitem[Zaidman et al.(2011)]{zaidman2011} Zaidman A, Van Rompaey B, van Deursen A, Demeyer S (2011) Studying the co-evolution of production and test code in open source and industrial developer test processes through repository mining. Empirical Software Engineering 16(3):325--364. \url{https://doi.org/10.1007/s10664-010-9143-7}

\bibitem[Zheng et al.(2023)]{zheng2023} Zheng L, Chiang WL, Sheng Y, Zhuang S, Wu Z, Zhuang Y, Lin Z, Li Z, Li D, Xing EP, Zhang H, Gonzalez JE, Stoica I (2023) Judging LLM-as-a-judge with MT-bench and chatbot arena. In: Advances in Neural Information Processing Systems 36 (NeurIPS) Datasets and Benchmarks Track. arXiv:2306.05685

\bibitem[Ziegler et al.(2022)]{ziegler2022} Ziegler A, Kalliamvakou E, Li XA, Rice A, Rifkin D, Simister S, Sittampalam G, Aftandilian E (2022) Productivity assessment of neural code completion. In: Proceedings of the 6th ACM SIGPLAN International Symposium on Machine Programming (MAPS), pp 21--29. \url{https://doi.org/10.1145/3520312.3534864}

\end{thebibliography}
\end{document}